\documentclass[12pt]{scrartcl}

\usepackage{graphicx}
\usepackage{makecell}
\usepackage{overpic}
\usepackage{tablefootnote}
\usepackage{threeparttable}
\usepackage{hyperref}
\usepackage{xurl}
\usepackage{breakurl}
\usepackage{tikz}
\usepackage{caption}
\usepackage{epstopdf}
\usepackage{subfigure}
\usepackage{float}
\usepackage{amsmath}
\usepackage{setspace}
\usepackage[margin=1in]{geometry}
\usepackage{breakurl}

\usepackage[format=plain]{caption}

\title{Uncertainty Quantification in Portfolio Temperature Alignment}

\author{
    \normalsize Hendrik Weichel\thanks{Sustainable Finance Research Lab, Frankfurt University of Applied Sciences, Frankfurt, Germany; University of Huddersfield, Huddersfield, United Kingdom \texttt{hendrik.weichel@fb2.fra-uas.de}} \and
    \normalsize Aleksandr Zinovev\thanks{\texttt{aleksandr.k.zinovev@outlook.com}}\and 
    \normalsize Heikki Haario \thanks{Computational Engineering, LUT University, Lappeenranta, Finland, \texttt{heikki.haario@lut.fi}} \and
   \normalsize Martin Simon\thanks{Sustainable Finance Research Lab, Frankfurt University of Applied Sciences, Frankfurt, Germany, \texttt{martin.simon@fb2.fra-uas.de}}
}

\date{\normalsize\today}

\begin{document}
\maketitle

%\doublespacing

%% Abstract
\begin{abstract}
\noindent\textbf{Abstract:}
We present a novel Bayesian framework for quantifying uncertainty in portfolio temperature alignment models, leveraging the X-Degree Compatibility (XDC) approach with the scientifically validated Finite Amplitude Impulse Response (FaIR) climate model. This framework significantly advances the widely adopted linear approaches that use the Transient Climate Response to Cumulative $\text{CO}_2$ Emissions (TCRE). Developed in collaboration with right°, one of the pioneering companies in portfolio temperature alignment, our methodology addresses key sources of uncertainty, including parameter variability and input emission data across diverse decarbonization pathways. By employing adaptive Markov Chain Monte Carlo (MCMC) methods, we provide robust parametric uncertainty quantification for the FaIR model. To enhance computational efficiency, we integrate a deep learning-based emulator, enabling near real-time simulations. Through practical examples, we demonstrate how this framework improves climate risk management and decision-making in portfolio construction by treating uncertainty as a critical feature rather than a constraint. Moreover, our approach identifies the primary sources of uncertainty, offering valuable insights for future research.\end{abstract}
%% Keywords
%\begin{keyword}
%Portfolio Temperature Alignment \sep Uncertainty Quantification \sep Climate Risk Management \sep Decarbonization Pathways 
%\end{keyword}

%% Main Text
\section{Introduction}

\noindent Transitioning to a net-zero society demands substantial capital investment to support the adoption of new low-emission technologies, particularly in high-emitting sectors. Financial institutions are pivotal in directing capital towards activities that align with a low-emission future. Rather than divesting, engaging with emission-intensive industries is recognized as a more effective strategy for driving this transition. Abandoning emission-heavy sectors like steel or cement is not a viable option, as their outputs will remain essential well beyond 2050. Therefore, fostering innovation within these sectors becomes imperative. Financial institutions must develop the capability to distinguish between companies that are leading the transition and those that are lagging behind. The potential of financial markets to catalyze corporate innovation has been emphasized, as highlighted, for example, in the work \cite{chang2019credit}.

From the risk management perspective, transition towards a decarbonized economy and the corresponding uncertainty may pose significant risks to both the credit and investment portfolios of financial institutions, encompassing credit, market, operational, and liquidity risks, see, e.g., \cite{bourgey2024bridging, bourgey2024efficient}. There is evidence that investors demand compensation for exposure to carbon emission risk and its associated uncertainties, see, e.g., \cite{bolton2021investors,avramov2022sustainable,pastor2021sustainable}, and that their expectations closely align with scientific projections, see \cite{schlenker2021market}. Companies in emission-intensive sectors need to optimize their business model adaptation plans under transition scenarios, cf.\cite{ndiaye2024optimal}, as they may face reduced competitiveness if they fail to adapt to the decarbonization process. Factors contributing to decreased competitiveness include higher carbon prices, regulatory changes, and evolving consumer preferences. These risks increase default probabilities, especially in scenarios where the transition is delayed, potentially leading to unforeseen losses for financial institutions heavily invested in such sectors. Given the reliance of many banks on interest income from emission-intensive sectors, these institutions in particular must thoroughly evaluate transition risks in their credit portfolios to mitigate potential losses. During the last years, portfolio alignment tools have emerged to facilitate this process, addressing challenges such as evaluating emissions relative to transition pathways, accounting for varied decarbonization rates, and projecting future transition performance. Portfolio alignment complements other techniques like scenario analysis and stress testing. While traditional metrics concentrate on the present-day carbon footprint of a portfolio, portfolio alignment metrics offer insight into the constituent companies' progress toward low-emission operations. The ECB's 2022 review on climate-related and environmental risks highlights the significance of alignment assessment in evaluating transition risks within credit risk management processes, cf. \cite{ECB2022,ECB2022_1}. The Portfolio Alignment Team's technical report \cite{Portfolio_alignment_team} aims to highlight emerging best practices in developing portfolio alignment tools; it states that \lq\lq portfolio alignment [...] will be subject to various sources of uncertainty arising from choice of methodology, data, and scenario\rq\rq\ and that \lq\lq portfolio managers should consider quantifying and disclosing the uncertainties associated with their portfolio or sub-portfolio alignment\rq\rq. Undeniably, the robustness of \emph{any} financial model depends on the accuracy and reliability of its output. Therefore, the present paper advocates for the adoption of systematic uncertainty quantification in portfolio alignment to enable consistent and robust portfolio alignment practices thus enhancing comparability, transparency, and clarity for all stakeholders involved. In this work, we focus on the newest category of portfolio alignment tools, known as \emph{implied temperature rise (ITR)} or \emph{temperature alignment models}. These models build on the foundation of traditional benchmark-divergence approaches \cite{Portfolio_alignment_team} but advance them by incorporating a physics-based climate projection. Most commonly, they rely on linear TCRE multiplier approaches, which yield a so-called \emph{implied temperature}. Roughly speaking, assigning an implied temperature of 2.5°C to a given portfolio indicates that the portfolio's constituent companies are surpassing their fair share of the global carbon budget in such a way that if everyone exceeded their fair shares by a similar proportion, the world would face approximately 2.5°C of mean global warming by the end of the century. Obviously, this statement depends, e.g., on the underlying modeling methodology. 

In the present work, we utilize a variant of the X-Degree Compatibility (XDC) model, first introduced by Helmke et al. \cite{helmke2020p} and the team at right°. The XDC model is a state-of-the-art temperature alignment tool, as evidenced by its recent application by the European Banking Authority (EBA) to assess banks’ alignment with the Paris Agreement temperature targets. Specifically, the EBA used the model to quantify the implied temperature rise of banks’ (non-SME) corporate loan books, leveraging granular exposure-level data collected from selected EU banks, see \cite{EBA2024}. The alignment cookbooks \cite{align, align2} provide a comprehensive overview of the different methodologies used by financial institutions, including the XDC model. All these models face multiple sources of uncertainty such as model, parameter, emission data and scenario uncertainty. While scenario uncertainty in climate risk management has been extensively investigated in several recent scientific publications \cite{le2022corporate, flora2023green, dumitrescu2024energy}, to the best of the authors' knowledge, there are no publications so far that include such approaches into temperature alignment methods. Although parameter uncertainties in simple climate models, such as the FaIR model, see \cite{smith2018fair}, and uncertainty in integrated assessment models, see, e.g., \cite{gillingham2018modeling,huard2022estimating}, have been previously studied using sensitivity-based methods, the corresponding frameworks have not been integrated with portfolio alignment tools either. A recent advancement in the temperature alignment literature is presented in \cite{implied}, which introduces a sensitivity-based approach for a linearized TCRE-based model. The work \cite{raynaud2020portfolio} highlights key uncertainties in temperature alignment, including those related to the distribution of Scope 3 emissions, avoided and removed emissions along the value chain, various types of scenario-related uncertainties, and the non-linearity of the climate response. We identify the lack of robust methodologies to address these uncertainties as a critical gap in both academic literature and the market for portfolio alignment tools -- a gap which we address in this work.

To the best of the authors' knowledge, the present work is the first to provide comprehensive uncertainty quantification -- including model, parameter, emission data, and scenario uncertainty -- for portfolio temperature alignment. This addresses the aforementioned research gap through a twofold contribution: First, we propose a methodological framework that utilizes the FaIR simple climate model rather than a linearized TCRE-based approximation, integrating all identified sources of uncertainty while delivering reduced uncertainty estimates compared to existing sensitivity-based approaches. Second, despite employing state-of-the-art adaptive Markov Chain Monte Carlo (MCMC) sampling algorithms, our framework maintains computational efficiency through the incorporation of a deep learning surrogate for the MCMC sampler, making it suitable for real-time portfolio analysis and trading desk applications. Supplementary code is available on GitHub\footnote{ \url{https://github.com/hendrikkwe/Uncertainty_in_Temperature_Alignment}}, facilitating reproducibility and broader applicability of the framework. 

The rest of the paper is structured as follows: Section 2 introduces our portfolio temperature alignment methodology. Section 3 discusses the various sources of uncertainty. Section 4 presents our framework and the underlying methodology, while Section 5 provides example computations. Finally, Section 6 concludes the paper.

%%%%%%%%%%%%%%%%%%%%%%%%%%%%%%%%%%%%%%%%%%%%%%%%%%%%%%%%%%%%%%%%%%%%%%%%%%%%%%%%%%%%%%%%%%%%

\section{Temperature alignment in financial climate risk management}
Temperature alignment embraces the idea of considering a forward-looking emissions pathway aligned with climate targets, rather than merely focusing on the present-day GHG emission footprint. 
The rationale behind this approach is that in order to achieve the temperature goal of the Paris Agreement, not all industry sectors must decarbonize at the same rate. Financial institutions must accurately account for this variability in their transition assessments which in turn necessitates making assumptions about the division of the global carbon budget across geography and sectors, as global warming depends on cumulative emissions rather than the emissions of individual actors. To enable such assessments, financial institutions require projections of how a counterparty's or portfolio's transition performance will evolve in the future. Once such projections are available, proactive engagement allows institutions to help counteract deviations from the required performance and steer the transition towards alignment with climate goals. Summarizing, measuring how a counterparty aligns with a specific global warming outcome requires three types of information: 
\begin{enumerate}
\item present-day GHG emission data 
\item forward-looking projections of future emissions (decarbonization plans)
\item a normative benchmark outlining the decarbonization pathway the counterparty needs to follow to achieve a specified climate target.
\end{enumerate} 
The availability of reliable present-day GHG emission data and decarbonization plans is one of the main challenges of such a forward-looking approach. Data gaps, consistency and comparability issues remain a key challenge and therefore a major source of uncertainty. Counterparty-level present-day GHG emission data is still not readily available, particularly for smaller and non-listed corporate counterparties. Even for large companies, measuring, tracking and reporting Scope 3 emission data defined by the GHG Protocol presents challenges. Looking into the 2022 ECB stress test results, the majority of participants used estimates of GHG emissions. And many only used Scope 1 and Scope 2 emission data. Reporting of counterparty-level decarbonization plans is still in its infancy and should gear up soon as historical performance will be crucial for assessing the credibility of future plans. In January 2023, the Corporate Sustainability Reporting Directive (CSRD), which applies to an estimated number of 50.000 companies in Europe, entered into force and it addresses exactly the need for better data. Notably, CSRD regulation not only aims to provide standards for transparent reporting, enabling informed investment and lending decisions, it also prescribes a science based temperature alignment of both investment and credit portfolios. Normative benchmarks are provided by forward-looking climate scenarios like those in the International Institute for Applied Systems Analysis Shared Socioeconomic Pathways (IIASA SSP) scenario database or the International Energy Agency's (IEA) projections. These scenarios, derived from climate-economy Integrated Assessment Models, outline specific pathways for emissions reduction and production capacity evolution. They provide insights into how different sectors of the economy must adjust to meet a specific climate target, considering various socioeconomic factors and delineate potential divisions of the global carbon budget across time, geography, and sectors to limit warming. SSP scenarios are gaining increasing importance in the financial climate risk sphere, see, e.g., \cite{bourgey2024bridging,bourgey2024efficient}, most notably as the foundation for the NGFS scenarios \cite{bertram2020ngfs}. There is a number of temperature alignment tools available in the market and these tools range in complexity, see, e.g., \cite{align,align2}. 

\subsection{The X-Degree Compatibility (XDC) model} \label{Sec:XDC}
The XDC model which was first proposed in \cite{helmke2020p} is among the pioneering models linking economic projections and physical climate modeling. The calculation is based on carefully predicting the answer to the question: How many emissions will a counterparty or portfolio cause in order to generate 1 million EUR of Gross Value Added (GVA) between today and 2050. With the evolution of complex benchmark-divergence models that leverage forward-looking climate scenarios to break down the global carbon budget into region- and sector-specific benchmarks, the XDC model enables portfolio managers to assess alignment with an \lq\lq X°C global warming\rq\rq\ pathway. A key strength of the XDC model is its full transparency and peer-reviewed foundation, which may have influenced the EBA's decision to adopt it for their recent assessments \cite{EBA2024}.

In this work, we provide a modification of the original XDC model which is particularly suited for portfolio construction: Namely we dissect the worldwide industry into industrial sectors which are weighted according to their present-day emissions, e.g., the electricity and heat sector is the largest contributor of GHG emissions, followed by the transport sector, etc. For each of these sectors we take into account science based decarbonization pathways.
In other words, we look at the worldwide production of GHGs as it currently stands, across all sectors. This contrasts with the construction methodology of so-called EU Paris-aligned benchmarks (PABs), see, e.g., \cite{wang2021eu}, which incentivize reallocating capital to low-emission sectors, thereby risking a decoupling of the benchmarks from the real economy.
 
\subsubsection{The socio-economic model}\label{subsubsec:soc-eco-model}
In the initial stage, the variant of the XDC model which we propose here establishes the Economic Emission Intensity (EEI) for each of the companies in the portfolio under consideration as the ratio of GHG emissions (measured in tonnes $\text{CO}_2$eq) to every million EUR of gross value added (GVA), where GVA is defined as the sum of Earnings Before Interest, Taxes, Depreciation, and Amortization (EBITDA) and personnel costs. The authors of the original XDC model opt for GVA as the benchmark because it represents the genuine value that a company generates between expenses and revenues, free from distortions caused by taxation and interest rates, cf. \cite{helmke2020p}. The present-day EEI of each of the portfolio constituents is computed for the base-year:
\begin{equation*}
\text{EEI}^{\text{Company}}=\frac{\text{EMISSIONS}^{\text{Company}}}{\text{GVA}^{\text{Company}}}.
\end{equation*} 

In the second step, we compute a weighted average $\text{EEI}^{\text{Portfolio,Sector}}$ of these EEIs for all of the portfolio constituents which belong to the same sector, weighted according to their absolute GHG emissions.
Each of the $\text{EEI}^{\text{Portfolio,Sector}}$ portfolio-sector intensities is then upscaled to a portfolio-specific, global, present-day sector emission value
\begin{equation*}
\text{EMISSIONS}^{\text{Portfolio,Sector}} = \text{EEI}^{\text{Portfolio, Sector}}\text{GVA}^{\text{Sector}}
\end{equation*}
thus approximately answering the question: \emph{Which amount of GHG emissions would be released into the atmosphere during the baseline-year, if the whole sector would operate as emission-intensively as the companies within the sector in the portfolio under consideration do?} This procedure is carried out for each of the sectors represented in the portfolio. For those sectors not represented in the portfolio, we stick with the original data for the sector.

In the third step, as a benchmark, we use a variety of science based socio-economic pathways from the SSP framework \cite{o2014new} to project the present-day sector emission performances into the future. This is done using the (simple) climate model described in the next subsection. A particularly valuable feature of the SSP framework is that each scenario is accompanied by a narrative describing the respective future world, cf. \cite{o2017roads}.
Given the availability of reliable counterparty-specific decarbonization pathways, this benchmark should subsequently be compared to a portfolio temperature derived from these pathways. Assessing the quality of decarbonization pathways reported by companies is a highly complex task that requires deep domain expertise. Despite these challenges, we consider this evaluation essential for ensuring the credibility and effectiveness of climate strategies. Importantly, our framework is designed to seamlessly integrate the resulting uncertainties, providing a comprehensive approach to analyzing their implications. However, resolving these complexities goes beyond the scope of this work. For now, we limit ourselves to benchmark calculations based on counterparty-specific present-day GHG emission data, leaving the integration of both, future counterparty-specific pathways and the corresponding uncertainty for future research.

\subsubsection{The climate model} \label{subsubsec:climate_model}
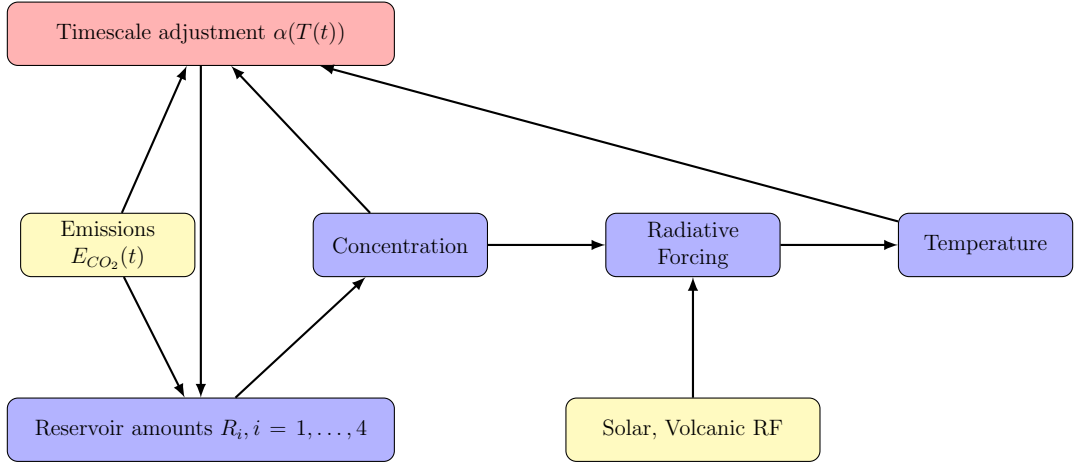
\begin{figure}[t!]
    \centering
    \begin{tikzpicture}[node distance=2.5cm, scale=0.7, transform shape]

        % Nodes
        \node (emissions) [rectangle, draw, fill=yellow!30, text width=3cm, text centered, rounded corners, minimum height=1.2cm] {Emissions $E_{CO_2}(t)$};
        \node (reservoir) [rectangle, below of=emissions, draw, fill=blue!30, text width=7cm, text centered, rounded corners, minimum height=1.2cm, yshift=-1cm, xshift=1.75cm] {Reservoir amounts $R_i, i = 1, \ldots, 4$};
        \node (concentration) [rectangle, right of=emissions, draw, fill=blue!30, text width=3cm, text centered, rounded corners, minimum height=1.2cm, xshift=3cm] {Concentration};
        \node (radiative) [rectangle, right of=concentration, draw, fill=blue!30, text width=3cm, text centered, rounded corners, minimum height=1.2cm, xshift=3cm] {Radiative Forcing};
        \node (temperature) [rectangle, right of=radiative, draw, fill=blue!30, text width=3cm, text centered, rounded corners, minimum height=1.2cm, xshift=3cm] {Temperature};
        \node (timescale) [rectangle, above of=concentration, draw, fill=red!30, text width=7cm, text centered, rounded corners, minimum height=1.2cm, yshift=1.5cm, xshift=-3.75cm] {Timescale adjustment $\alpha(T(t))$};
        \node (solar) [rectangle, below of=radiative, draw, fill=yellow!30, text width=4.5cm, text centered, rounded corners, minimum height=1.2cm, yshift=-1cm] {Solar, Volcanic RF};

        % Arrows
        \draw[->, >=latex, thick] (reservoir) -- (concentration);
        \draw[->, >=latex, thick] (concentration) -- (radiative);
        \draw[->, >=latex, thick] (radiative) -- (temperature);
        \draw[->, >=latex, thick] (temperature) -- (timescale);
        \draw[->, >=latex, thick] (timescale) -- (reservoir);
        \draw[->, >=latex, thick] (concentration) -- (timescale);
        \draw[->, >=latex, thick] (solar) -- (radiative);
        \draw[->, >=latex, thick] (emissions) -- (reservoir);
        \draw[->, >=latex, thick] (emissions) -- (timescale);

    \end{tikzpicture}
    \caption{Schematic workflow of the FaIR simple climate model computation, cf. \cite{smith2018fair}.}
    \label{fig:FaIR_model}
\end{figure}

In this work, we use the FaIR simple climate model rather than a linear TCRE approach which suffers from a number of limitations:
Some studies find the TCRE is not fully path-independent as it varies slightly with the rate of $\text{CO}_2$ emissions, see, e.g., \cite{krasting2014trajectory}. Moreover, the approximate linear relationship only holds for $\text{CO}_2$, it does not account for warming from other greenhouse gases such as methane, cf. \cite{dietz2019cumulative}. A key feature in earth system models is a positive carbon cycle feedback, meaning that as surface temperature increases, land and ocean carbon sinks become less effective at absorbing $\text{CO}_2$ and a larger proportion of any further emitted carbon will remain in the atmosphere. The Finite Amplitude Impulse Response (FaIR) model, first introduced in \cite{millar2017modified} has since undergone several modifications (see \cite{smith2018fair,leach2021fairv2,smith2024fair}). In this work, we follow the exposition provided in \cite{smith2018fair}. The FaIR model computes the global temperature and the atmospheric GHG concentrations in steps of one year. It takes the previous years concentrations and the yearly emissions of greenhouse gases as its main input and determines the atmospheric concentrations of greenhouse gases taking into account the decay in the atmosphere due to sinks (reservoirs) in the carbon cycle, namely geological, deep ocean, biosphere and the ocean mixed layer. Figure \ref{fig:FaIR_model} provides a workflow diagram illustrating the steps and processes involved: The decay in these reservoirs happens over different timespans depending on both, the reservoir and the temperature. The governing equation for the amount $R_i$ of $\text{CO}_2$ in each reservoir is
\begin{equation*}
\frac{\mathrm d R_i}{\mathrm{d} t}=a_iE_{\text{$\text{CO}_2$}}(t)-\frac{R_i}{\alpha(T(t))\tau_i},\quad i=1,...,4,
\end{equation*}
where $E_{\text{$\text{CO}_2$}}(t)$ denotes the emission at time $t$; $a_i, i=1,...,4,$ the reservoir fractions ($\sum_{i=1}^4 a_i=1$); $\tau_i, i=1,...,4,$ the lifespan of $\text{CO}_2$ in reservoir $i$; and $\alpha(\cdot)$ the timescale adjustment for the temperature dependence of the lifespan.
FaIR also calculates non-$\text{CO}_2$ greenhouse gas concentrations from emissions, aerosol forcing from aerosol precursor emissions, tropospheric and stratospheric ozone forcing from the emissions of precursors, and forcings from black carbon on snow, stratospheric methane oxidation to water vapour, contrails and land use change. Forcings from volcanic eruptions and solar irradiance fluctuations are supplied externally. These forcings are then converted to a temperature change. 
The impacts of different forcing factors are tuned  so that FaIR simulations agree with the results of more detailed large climate models.
The various forcing terms are modelled as parametrized formulas. This leads to a relatively high number of parameters that need to be estimated, thereby resulting in increased parameter uncertainty. We will discuss this issue in Section 3 below.

%%%%%%%%%%%%%%%%%%%%%%%%%%%%%%%%%%%%%%%%%%%%%%%%%%%%%%%%%%%%%%%%%%%%%%%%%%%%%%%%%%%%%%%%%%%%

\section{Sources of uncertainty}
In this section, we examine the various sources of uncertainty included in our framework. Understanding these sources is crucial for accurately modeling their impact and improving the robustness of our analysis.

\subsection{Parameter uncertainty}
The FaIR model has 20 parameters which need to be calibrated to historical temperature and concentration data. This calibration problem is ill-posed in the sense that given the limited amount of data, the solution to the problem will be non-unique. Attempts to estimate all parameters without regularization typically lead to parameter values which are not compatible with existing physics domain expertise. Even when using physics-based priors for the parameters as a means to regularize the problem, going with one particular calibration of these parameters (obtained via optimization) introduces errors, as it will not capture the full variability in the model outputs which poses the risk of introducing a \lq\lq hot\rq\rq\ or \lq\lq cold\rq\rq\ bias. Therefore, we consider quantification of the inherent parameter uncertainty a crucial ingredient to guarantee prudent risk management.
In Subsection \ref{sec:UQ} below, we describe our framework to quantify parameter uncertainty in a Bayesian setting.

\subsection{Emission data uncertainty}
Estimating present-day GHG emissions is a challenging task, cf. \cite{Unc_edgar} and therefore emission data uncertainty needs to be modeled and propagated through temperature alignment models. While modeling and estimating present-day GHG emission uncertainty is beyond the scope of this work, we demonstrate in Subsection \ref{subsec:test_emissions_unc} below, how emission data uncertainty can be integrated into our framework.

\subsection{Model uncertainty}
Climate models range from simple, conceptual frameworks to complex, highly detailed simulations. While the so-called \emph{simple} climate models are the way to go for portfolio temperature alignment because of their ease of use and ability to provide quick insights, they come with significant model uncertainties that must be carefully considered. This inevitable structural uncertainty arises from the simplifications and assumptions inherent in the models' design. Simple models usually omit or approximate processes that are critical in more complex systems. For instance, feedback mechanisms like ice-albedo feedback or water vapor feedback might be oversimplified, leading to incomplete or biased results. Different simple climate models may produce different results due to variations in their underlying assumptions, parameter choices, and structural designs. Therefore, we strongly recommend to address this model uncertainty by taking into account a number of simple climate models. Comparing outputs from multiple models can help identify the range of possible outcomes, but it also highlights the uncertainty stemming from model selection. In Subsection \ref{sec:ModelUQ} below, we describe how our framework can provide such a comparison in a Bayesian setting.

\subsection{Scenario uncertainty} \label{sec:scenario_uncertainty}
To determine an implied temperature increase from the FaIR climate model, a pathway of future annual economic emission intensities as well as an economic growth scenario are required as inputs. However, projecting a company's or sector's climate and economic performance into the future is a significant challenge akin to attempting to predict the future. Therefore, a comprehensive uncertainty analysis, taking into account a wide range of different climate and economic scenarios is a necessity. While it is important to develop a deep understanding of different scenarios from different sources among risk practitioners, we focus in this work on the so-called representative concentration pathways (RCPs) established in the IPCC AR5 \cite{stocker2014climate}. An RCP is a GHG concentration trajectory, the AR5 database\footnote{\url{https://tntcat.iiasa.ac.at/RcpDb/dsd?Action=htmlpage&page=welcome}} comprises 1184 scenarios generated using different integrated assessment models and among those, the RCP 1.9, RCP 2.6, RCP 3.4, RCP 4.5, RCP 6, RCP 7 and RCP 8.5 have been selected to represent reference scenarios. The RCPs are named according to their 2100 radiative forcing level as predicted by the individual models. These RCP scenarios have been replenished with the Shared Socioeconomic Pathways (SSPs) introduced by O'Neill et al. 2014 \cite{Neill_SSPs}. These socio-economic scenarios provide a framework for understanding different pathways society might take considering varying degrees of sustainability, cooperation and reliance on fossil fuels.
The five narratives which were initially developed in \cite{Neill_SSPs} to describe different societal developments are visualized in Figure \ref{SSP_MAP} below and summarized in the Appendix. In contrast to the RCP scenarios, the SSP-based scenarios provide an economic and societal rationale for the assumed emission trajectories and land use changes. Additionally, they incorporate updated historical emissions of greenhouse gases and aerosols, as well as land use changes. For practical scenario uncertainty quantification, we recommend to take each of the following SSP-RCP mappings into account: SSP1-RCP1.9, SSP1-RCP2.6, SSP2-RCP4.5, SSP3-RCP7.0 and SSP5-RCP8.5.

\begin{figure}[t!]
\centering
\begin{tikzpicture}[scale=2]
    % Define the colors
    \definecolor{ssp1}{RGB}{144,238,144} % Light green
    \definecolor{ssp2}{RGB}{135,206,250} % Light blue
    \definecolor{ssp3}{RGB}{173,216,230} % Light purple
    \definecolor{ssp4}{RGB}{255,222,173} % Light orange
    \definecolor{ssp5}{RGB}{255,160,122} % Light red
    
    % Draw the squares
    \fill[ssp1] (0,0) rectangle (2,2);
    \fill[ssp5] (0,2) rectangle (2,4);
    \fill[ssp3] (2,2) rectangle (4,4);
    \fill[ssp4] (2,0) rectangle (4,2);
    
    % Draw the diamond
    \fill[ssp2] (1,2) -- (2,3) -- (3,2) -- (2,1) -- cycle;
    
    % Draw the arrows for the axes
    \draw[->, thick] (-0.5,0) -- (4.5,0) node[below, midway] {Challenges for adaptation};
    \draw[->, thick] (0,-0.5) -- (0,4.5) node[above, midway, sloped] {Challenges for mitigation};    
    % Add the labels with increased font size
    \node[align=center, font=\large] at (1,3) {\textbf{SSP5}\\ Fossil-fueled dev.};
    \node[align=center, font=\large] at (3,3) {\textbf{SSP3}\\ Regional rivalry};
    \node[align=center, font=\large] at (3,1) {\textbf{SSP4}\\ Inequality};
    \node[align=center, font=\large] at (1,1) {\textbf{SSP1}\\ Sustainability};
    \node[align=center, font=\large] at (2,2) {\textbf{SSP2}\\ Middle of the road};
\end{tikzpicture}
\caption{Representation of SSP scenarios with challenges for adaptation and mitigation, adapted from \cite{Neill_SSPs}.}
\label{SSP_MAP}
\end{figure}
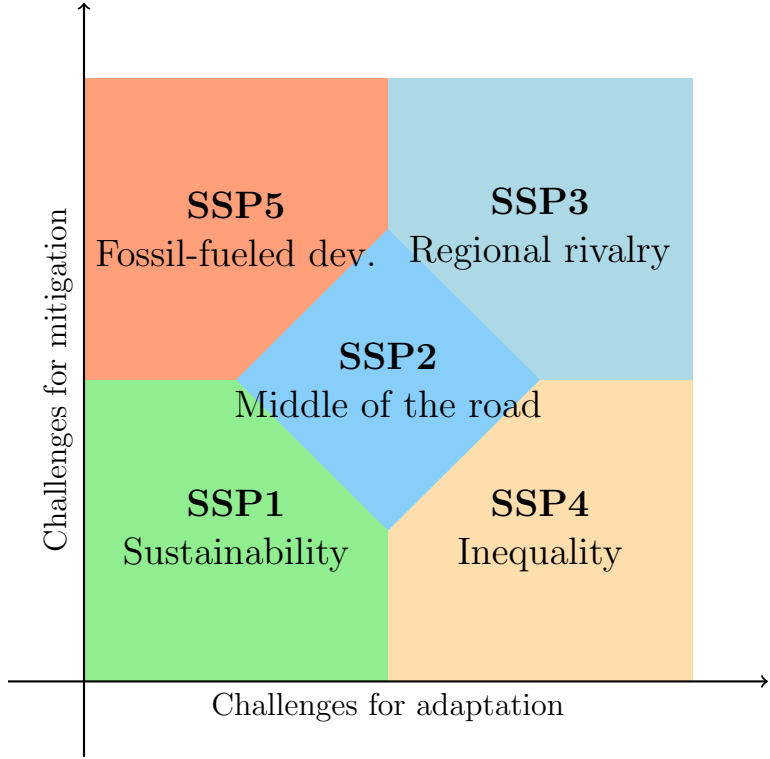

%%%%%%%%%%%%%%%%%%%%%%%%%%%%%%%%%%%%%%%%%%%%%%%%%%%%%%%%%%%%%%%%%%%%%%%%%%%%%%%%%%%%%%%%%%%%

\section{Methodology}

Assessing uncertainties is a critical step in portfolio temperature alignment. In our framework, we employ a Bayesian methodology to quantify parameter uncertainties in the FaIR climate model, resulting in posterior distributions that reflect the impact of observed data and prior knowledge. Emission uncertainties are incorporated into these distributions through probability density functions, allowing for a comprehensive representation of variability in reported data. Additionally, we propose a methodology for assessing model uncertainties, focusing on the structural and parametric assumptions inherent in the FaIR model. While this approach lays the groundwork, future research will be essential to implement and validate the methodology through experimental comparisons and scenario analyses.

\subsection{Bayesian parameter uncertainty analysis}\label{sec:UQ}
The parameter calibration problem for the FaIR model suffers from non-uniqueness: To fit the model to agree with historical records of global temperature, we have to tune 20 model parameters. Moreover, the knowledge of past temperatures is uncertain, all parameters that fit the model within the temperature error bars are acceptable. As a consequence,  there may be large ranges of possible (and physically reasonable) parameter vectors that lead to a sufficiently good fit to the observed historical temperatures. We address this issue by adopting a \emph{Bayesian} perspective, which offers two advantages: First, it enables the ability to incorporate prior knowledge or assumptions regarding the probability distribution of the unknown parameters. This is mostly knowledge of experts with domain expertise regarding where the solution of the inverse calibration problem is thought to be located. Second, this Bayesian approach produces more than merely point estimates, in fact it yields a \emph{posterior probability distribution} of possible solutions, each having a certain probability given the observed data, thus enabling uncertainty quantification. 

Let us therefore formulate the inverse parameter calibration problem in a statistical setting by writing our model as
\begin{equation}\label{eqn:sm}
\textbf{y}=F(\textbf{x}|\boldsymbol{\theta})+\boldsymbol{\varepsilon},
\end{equation}
where the vector $\textbf{y}$ contains observed measurements, i.e., the temperature and concentration measurements, $F$ denotes the mapping of the input emission vector $\textbf{x}$ to the measurements provided by the FaIR model. $F$ does neither take into account measurement noise nor modelling bias. Therefore, we model these by the additive noise vector
$\boldsymbol{\varepsilon}$. Finally, the vector $\boldsymbol{\theta}$ contains the model parameters that have to be estimated from the observed data. Given the statistical model (\ref{eqn:sm}), we apply \emph{Bayes’ Theorem} to obtain a mathematical expression for the corresponding \emph{posterior probability density} which can be explored numerically to quantify the uncertainties. We refer to Section \ref{ap:bayesian_posterior} of the Appendix for further details.

Sampling the posterior can be done using Markov Chain Monte Carlo (MCMC) methods, which offer a way to iteratively draw a chain of realizations of the random parameter vector, $\boldsymbol{\theta}$, from the posterior. In this work, we use Delayed Adaptive Metropolis (DRAM) \cite{haario2006dram}. DRAM is well-suited for the problem at hand as it combines the advantages of Adaptive Metropolis (AM), cf. \cite{haario2001adaptive} and Delayed Rejection (DR), cf. \cite{mira2001metropolis}. While the AM algorithm adapts the proposal distribution  based on the past history of the chain, the DR algorithm improves the efficiency of the resulting MCMC estimator by reducing the number of rejected proposals. In other words, AM allows for global adaptation of the proposal distribution based on all previously accepted proposals, while DR allows for local adaptation, only based on rejected proposals within each time-step. Once we have successfully sampled the posterior to calibrate the model to the past temperature and concentration observations, we can compute predictions by drawing from this posterior combined with independent draws from the distribution characterizing the emission data uncertainty using various decarbonization pathways.

\subsection{Input emission uncertainty analysis}\label{sec:EmissionUQ}

To incorporate uncertainties in present-day GHG emissions, we propose a flexible framework using suitable probability density functions (PDFs) to represent percentage deviations. Users can adjust the degree of uncertainty by modifying the parameters of these PDFs, according to their domain expertise. Random errors can be modeled using normal distributions, while systematic errors might be better captured using lognormal distributions.

Through Monte Carlo sampling, we propagate these uncertainties by modifying the entire emissions pathway based on an offset applied across the time series. This offset is determined as the product of the sampled error and the base-year emissions value. Iterating the FaIR model yearly with these adjusted pathways allows us to generate temperature trajectories that encapsulate the cumulative effect of input uncertainties over time. An example is provided in Test Case 2 below.

\subsection{Bayesian model uncertainty analysis}\label{sec:ModelUQ}
The basic idea of Bayesian model selection is as follows: Suppose that we have a set of possible (simple) climate models $\{\mathcal{M}_1,...,\mathcal{M}_k\}$, where model $\mathcal{M}_i$, $i=1,...,k$ possesses the likelihood density $\pi_i(\textbf{y}|\boldsymbol{\theta}_i)$ with model parameters $\boldsymbol{\theta}_i$ for the observed data $\textbf{y}$. Assigning prior probabilities to each model, we can compute the posterior probability $P(\mathcal{M}_i|\boldsymbol{y})$ of model $\mathcal{M}_i$, $i=1,..,k,$ given the data $\boldsymbol{y}$. Even though the set of models under consideration will certainly not contain the true data-generating model, Bayesian model uncertainty analysis will asymptotically assign probability one to the model closest to the true model with respect to the Kullback-Leibler divergence. In particular, although we leave the practical comparison of different climate models used by practitioners to future work, we would like to emphasize that our framework can provide a measure of model performance relative to other models. And even more, there is current research providing absolute goodness-of-fit assessments in comparing imperfect models, see, e.g., \cite{li2020} and the references therein. 

\subsection{Deep learning an emulator for model prediction including posterior uncertainty estimates}
One of the drawbacks of our framework, as described thus far, is its computational cost:
The run time for a single 80-year prediction using the FaIR model on a Macbook Pro with M2 Pro CPU is approximately 0.07 seconds. To quantify the parametric uncertainty of the model simulations for a given scenario, we repeatedly sample the posterior (the parameter chain produced by the MCMC run) around 10\,000 times to achieve the desired accuracy. This results in a run time of roughly 12 minutes per scenario, which may be prohibitive in practice, especially when near-realtime testing of multiple portfolios across a range of different scenarios is required.

To reduce computation time, we employ a machine learning emulator to accelerate our approach. 
The emulator is designed to accurately produce the same outputs as the MCMC sampling of the FaIR model, i.e., the posterior mean or median and confidence intervals for given scenario pathways and present-day GHG emissions. Although many neural network architectures are available for this task, training these networks in reasonable time is non-trivial: A brute-force approach to creating training data would be computationally intractable. For example, if minimum and maximum values were available for each of the 20 uncertain parameters of the FaIR model, and only a crude set of 5 linearly spaced values were used between those limits for each parameter, it would result in $5^{20}$ simulations--equivalent to roughly 500\,000 years of computing time. The use of the parameter posterior produced by MCMC solves this issue, as the posterior sampling \lq\lq automatically\lq\lq\ identifies the relevant parameter combinations, i.e., those that make the model fit the data statistically well enough. While MCMC sampling is also time-consuming, a sufficiently long chain (around one million samples) can be generated within one day on a standard machine.

The emulator takes emissions in the base year as input and calculates the projected mean earth temperature and 90\% credible interval under different scenarios (SSP1-RCP1.9, SSP1-RCP2.6, SSP2-RCP4.5, SSP3-RCP7.0, and SSP5-RCP8.5). For this task, we use a standard feed-forward neural network with three dense layers of 20 neurons. As a result, the run time is reduced from 12 minutes to 0.06 seconds. 

%%%%%%%%%%%%%%%%%%%%%%%%%%%%%%%%%%%%%%%%%%%%%%%%%%%%%%%%%%%%%%%%%%%%%%%%%%%%%%%%%%%%%%%%%%%%

\section{Numerical examples}
To demonstrate the feasibility of our approach and discuss its strengths and limitations, we present four test cases in this section. First, we apply the Bayesian methodology described in Subsection \ref{sec:UQ} for quantifying parameter uncertainty for an SSP scenario. The results highlight that this methodology significantly outperforms traditional sensitivity-based Monte Carlo simulations in terms of estimation uncertainty. In the second test case, we integrate emission uncertainty quantification as outlined in Subsection \ref{sec:EmissionUQ}, showcasing the importance of accounting for uncertainties in reported emission data. The third test case explores the combined application of parameter and emission uncertainty quantification to SSP scenarios. This demonstrates how their integration improves the overall analysis by capturing a wider range of uncertainty sources. Finally, we present a real-world example involving the Swedish steel producer SSAB AB. By applying our socio-economic model alongside the uncertainty quantification framework, we assess the climate-related transition risks faced by the company, providing a practical demonstration of the framework's utility in real-world decision-making. 

\subsection{Test Case 1: Comparison with sensitivity-based approach}
\begin{figure}[t!]
    \centering
    \includegraphics[width=\textwidth]{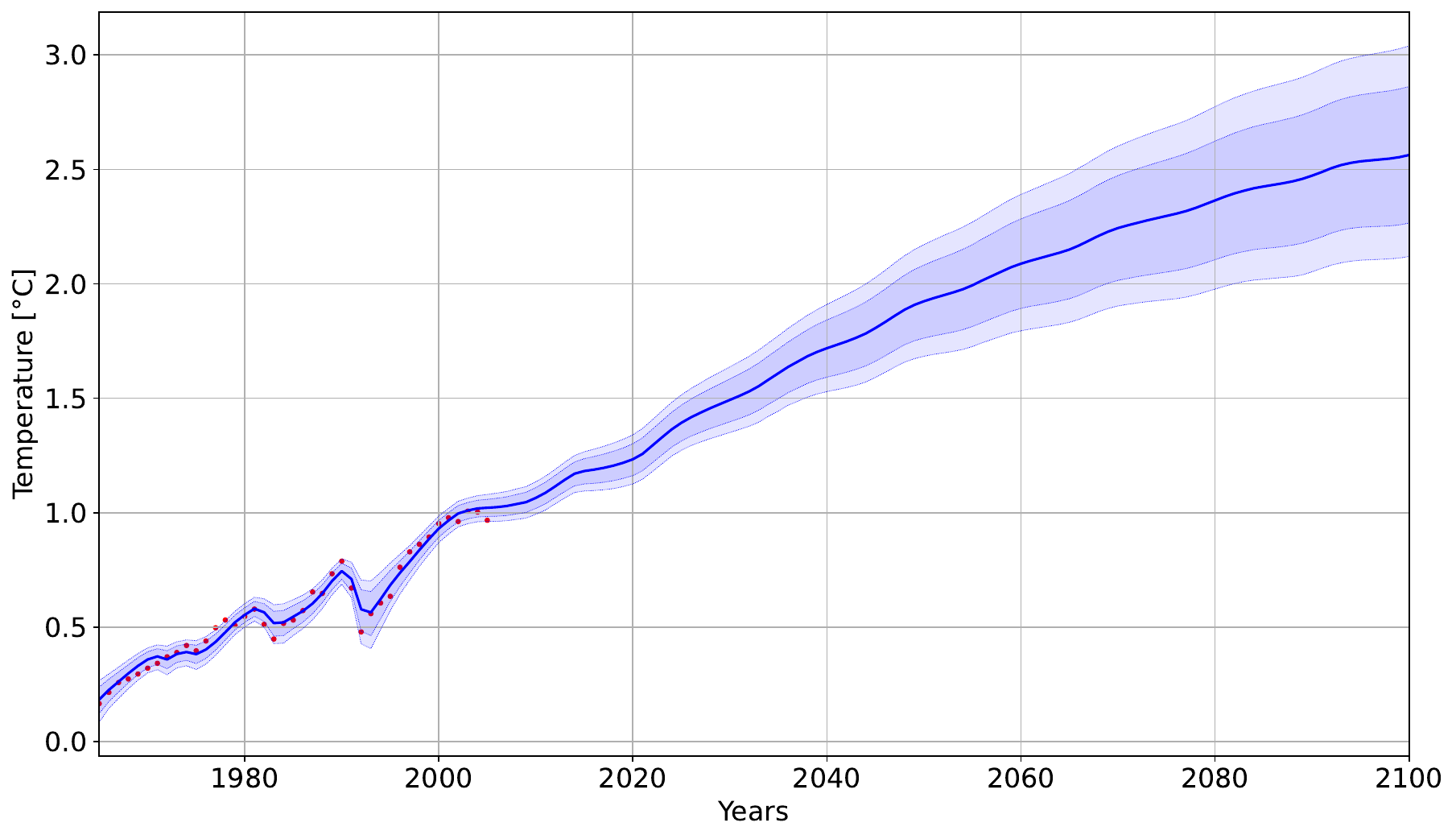}
    \captionsetup{font=footnotesize}
    \caption{Parameter UQ for SSP2-RCP4.5 Scenario, with Median (solid line) and 90\% and 99\% confidence. Red dots represent historical data.}
    \label{fig:temperature_parameter_uq_ssp2_rcp45}
\end{figure}

This test case exemplifies parameter uncertainty quantification. For parameter calibration we employed Bayesian modeling (see Section \ref{sec:UQ}) and applied Markov Chain Monte Carlo (MCMC) simulations to obtain the posterior distributions of FaIR's input parameters. A more detailed explanation is provided in Appendix Section \ref{ap:bayesian_posterior}. By sampling from these posterior distributions and running FaIR with emissions from the SSP2-RCP4.5 scenarios, we generated the temperature pathway shown in Figure \ref{fig:temperature_parameter_uq_ssp2_rcp45}, including 90\% and 99\% confidence intervals. In contrast, \cite{uq_Zinovev} conducted simple Monte Carlo sampling using the prior parameter distributions as outlined in the original FaIR paper \cite{smith2018fair}. Our approach, which integrates uncertainty quantification with MCMC parameter calibration, demonstrates a significant improvement in uncertainty estimation. Specifically, for the RCP-8.5 scenario the 90\% confidence interval for the temperature forecast in 2050 was (1.551, 3.971) using simple Monte Carlo sampling. In comparison, utilizing the posterior distributions from our MCMC calibration narrowed the 90\% confidence interval to (2.107, 2.484).

\subsection{Test Case 2: Comparison of different emission uncertainty models}
\label{subsec:test_emissions_unc}
This test case is inspired by recent discussions around the Science Based Target Initiative (SBTi), which has advocated for increased use of carbon credits in corporate net-zero plans \footnote{\sloppy\url{https://sciencebasedtargets.org/news/statement-from-the-sbti-board-of-trustees-on-use-of-environmental-attribute-certificates-including-but-not-limited-to-voluntary-carbon-markets-for-abatement-purposes-limited-to-scope-3}}. This advocacy lead to calls from SBTi employees for their CEO's resignation\footnote{\sloppy\url{https://www.theguardian.com/environment/2024/apr/11/climate-target-organisation-faces-staff-revolt-over-carbon-offsetting-plan-sbti}}. In 2023, an investigation by \emph{The Guardian} and Germany's \emph{ZEIT} titled \lq\lq Revealed: More than 90\% of Rainforest Carbon Offsets by Biggest Certifier Are Worthless\rq\rq\footnote{\sloppy\url{https://www.theguardian.com/environment/2023/jan/18/revealed-forest-carbon-offsets-biggest-provider-worthless-verra-aoe}} had highlighted significant uncertainties surrounding carbon offsets. Given these potentially substantial uncertainties, prudent risk management should incorporate these considerations into temperature alignment frameworks. 

\begin{figure}[H]
    \begin{minipage}{0.48\textwidth}
        \small
        \centering
        \includegraphics[height=4.4cm]{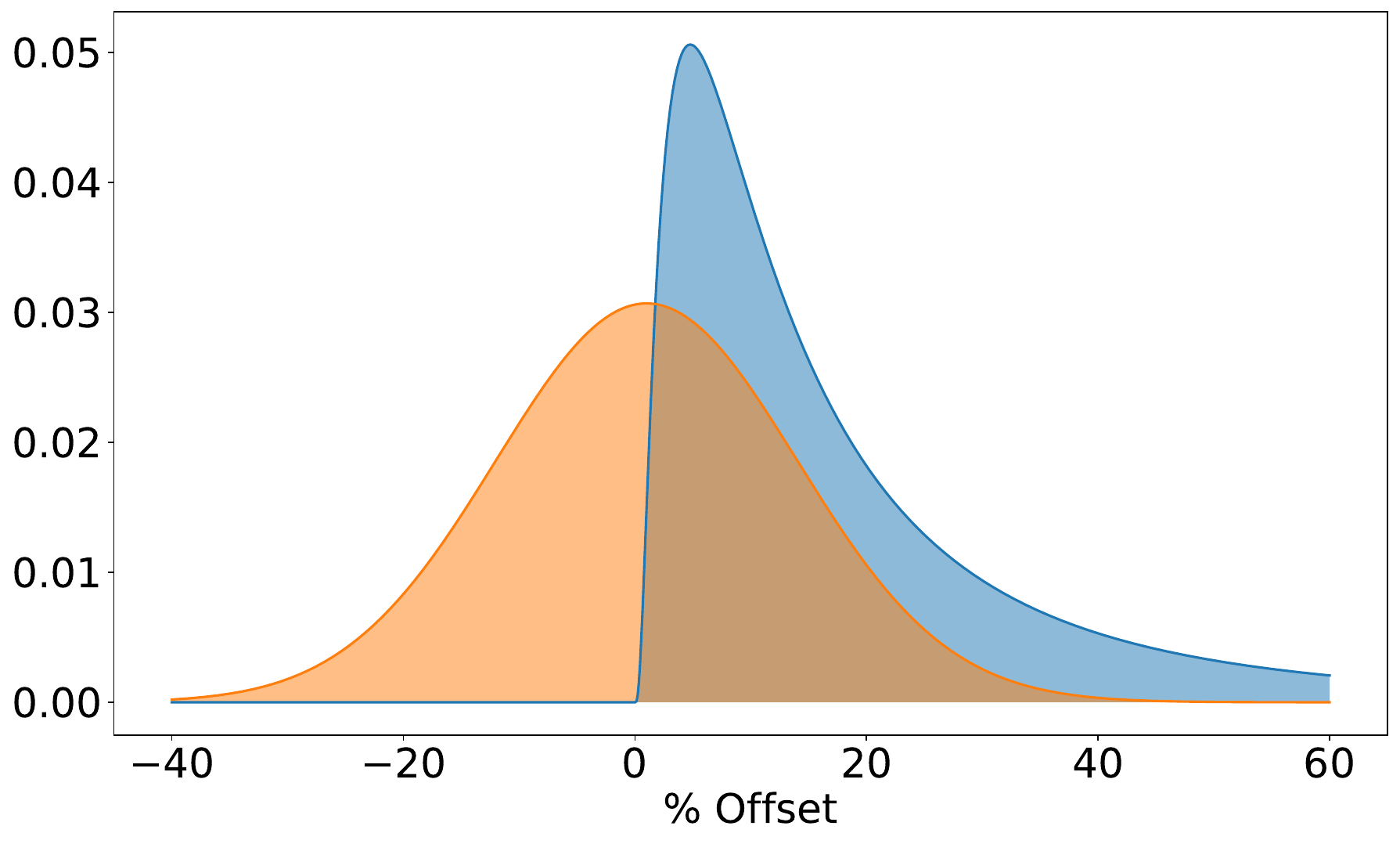}
        \captionsetup{font=footnotesize}
        \caption{PDFs from which the deviation in emissions is sampled. Blue lognormal distribution, Green normal distribution.}
        \label{fig:distribution_lognorm_norm}
    \end{minipage}
    \hfill
    \begin{minipage}{0.48\textwidth}
        \centering
        \includegraphics[height=4.4cm]{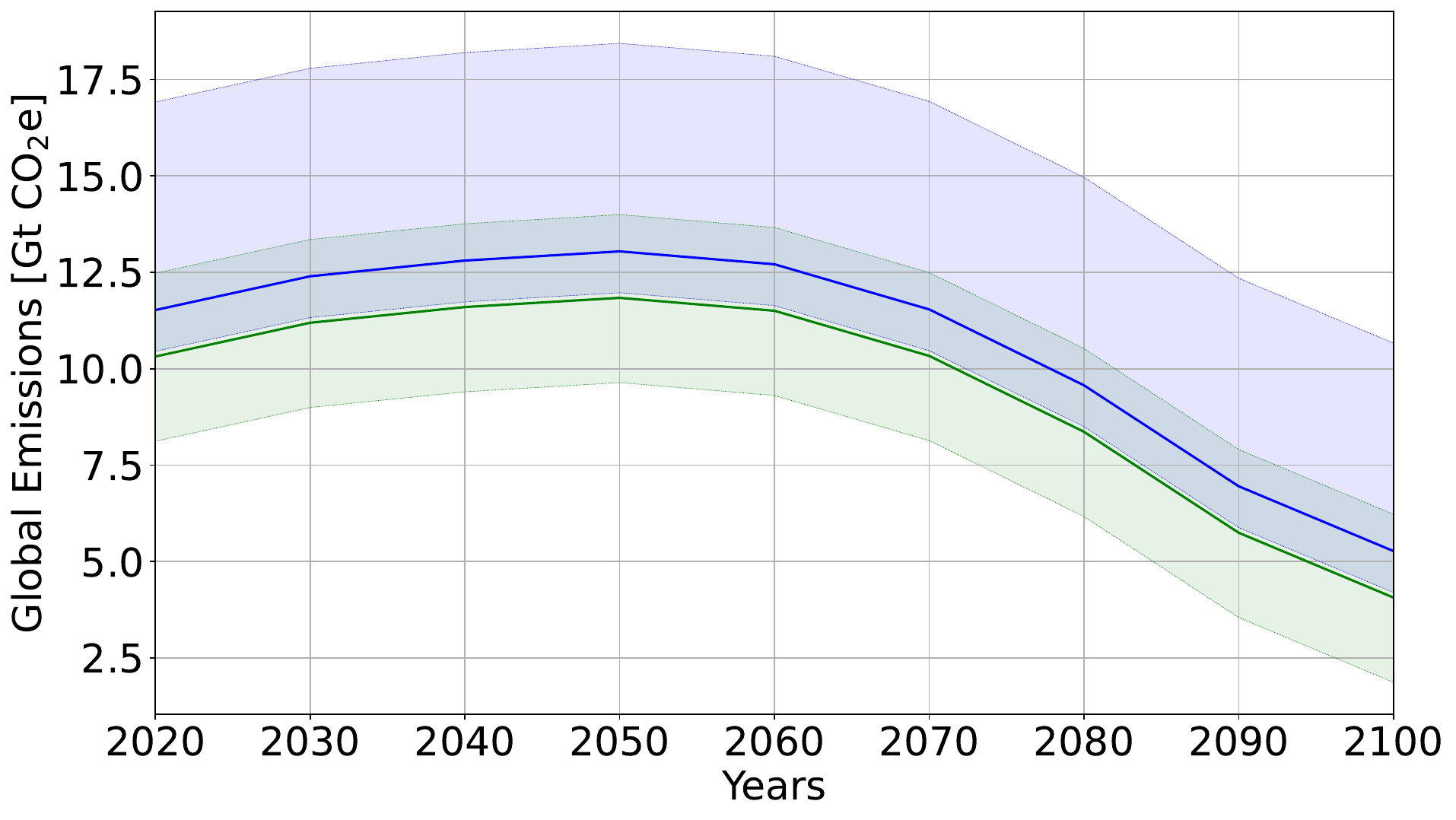}
        \captionsetup{font=footnotesize}
        \caption{90\% confidence interval of SSP2-RCP4.5 emission with lognormal (blue) and normal (green) uncertainty. The solid line is the median.}
        \label{fig:emissions_co2_uncertainty_lognorm}    
    \end{minipage}
\end{figure}

\begin{figure}[H]
    \centering
    \includegraphics[width=\textwidth]{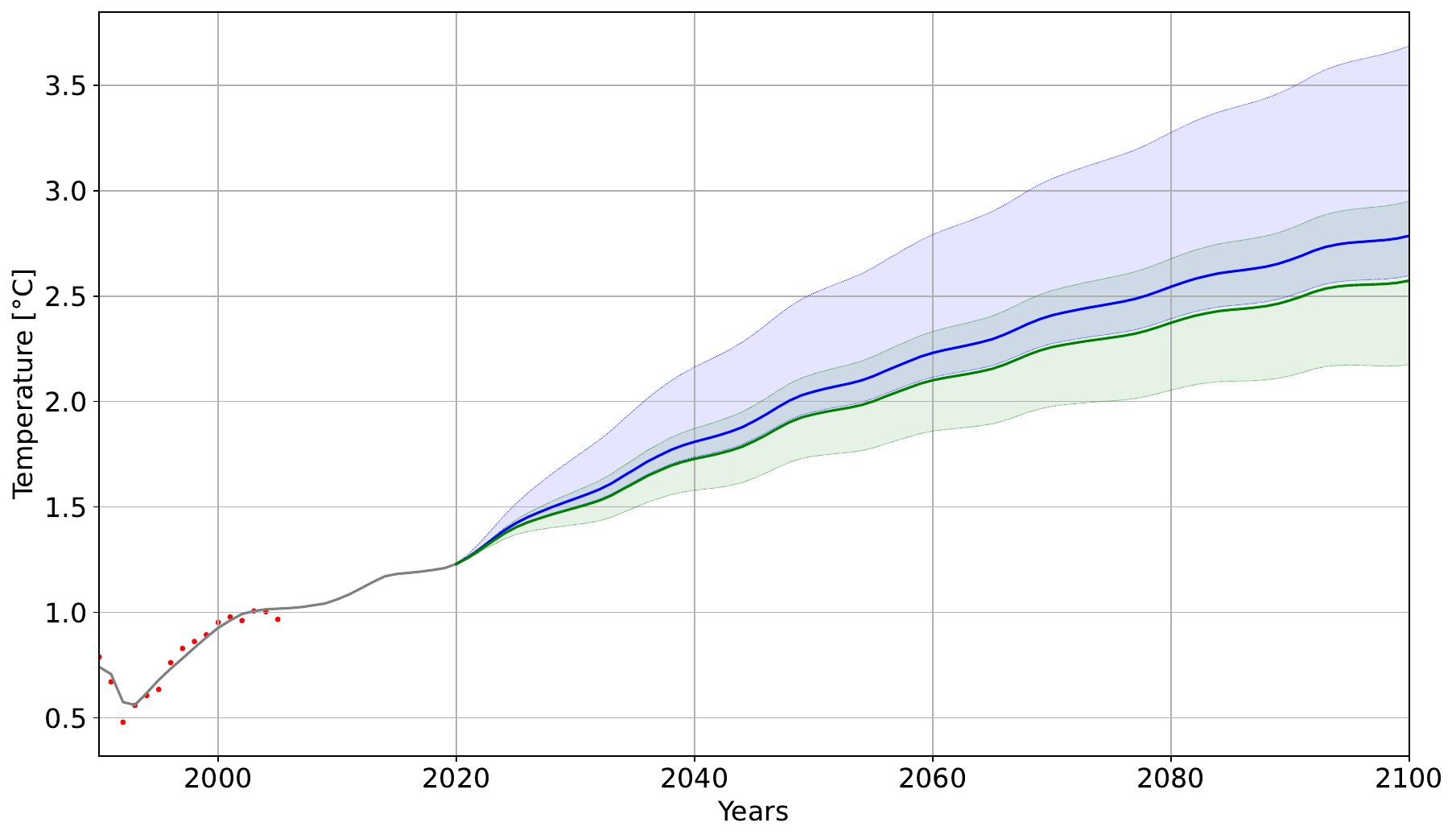}
    \captionsetup{font=footnotesize}
    \caption{90\% confidence interval for the implied temperature of SSP2-RCP4.5 scenario with emission uncertainty with lognormal and normal distribution ($\mu = 1$, $\sigma = 13$) applied to the pathways. The solid line is the median. Blue lognormal distribution, Green normal distribution. Red dots represent historical data.} 
    \label{fig:emissions_uq_temp_lognorm}
\end{figure}

We demonstrate the quantification of input emission uncertainties through modeling with probability density functions. Figure \ref{fig:distribution_lognorm_norm} presents the PDFs of lognormal and normal distributions, defined by parameters $\mu$ and $\sigma$, from which percentage deviations are sampled. The lognormal distribution is particularly suitable when there is evidence that reported emissions systematically underestimate true values, as may occur with the use of carbon offsets.
Figure \ref{fig:emissions_co2_uncertainty_lognorm} illustrates the emission pathway for the SSP2-RCP4.5 scenario, along with the 90\% confidence interval and median derived from the sampling process. Both distributions in this example use parameters $\mu = 1$ and $\sigma = 13$. In the final step, each sampled emission pathway is used to compute the corresponding temperature forecast, as shown in Figure \ref{fig:emissions_uq_temp_lognorm}. 

Future research should focus on refining these parameters. For example, \cite{EDGAR_UNC} investigates uncertainties in the Emissions Database for Global Atmospheric Research (EDGAR) emission inventory of greenhouse gases. Their findings indicate that \lq\lq the anthropogenic emissions covered by EDGAR for the combined three main GHGs for the year 2015 are accurate within an interval of -15\% to +20\% (defining the 95\% confidence of a log-normal distribution)\rq\rq.

\subsection{Test Case 3: Combined parameter and emission uncertainty quantification.}

\begin{figure}[H]
    \centering
    \includegraphics[width=\textwidth]{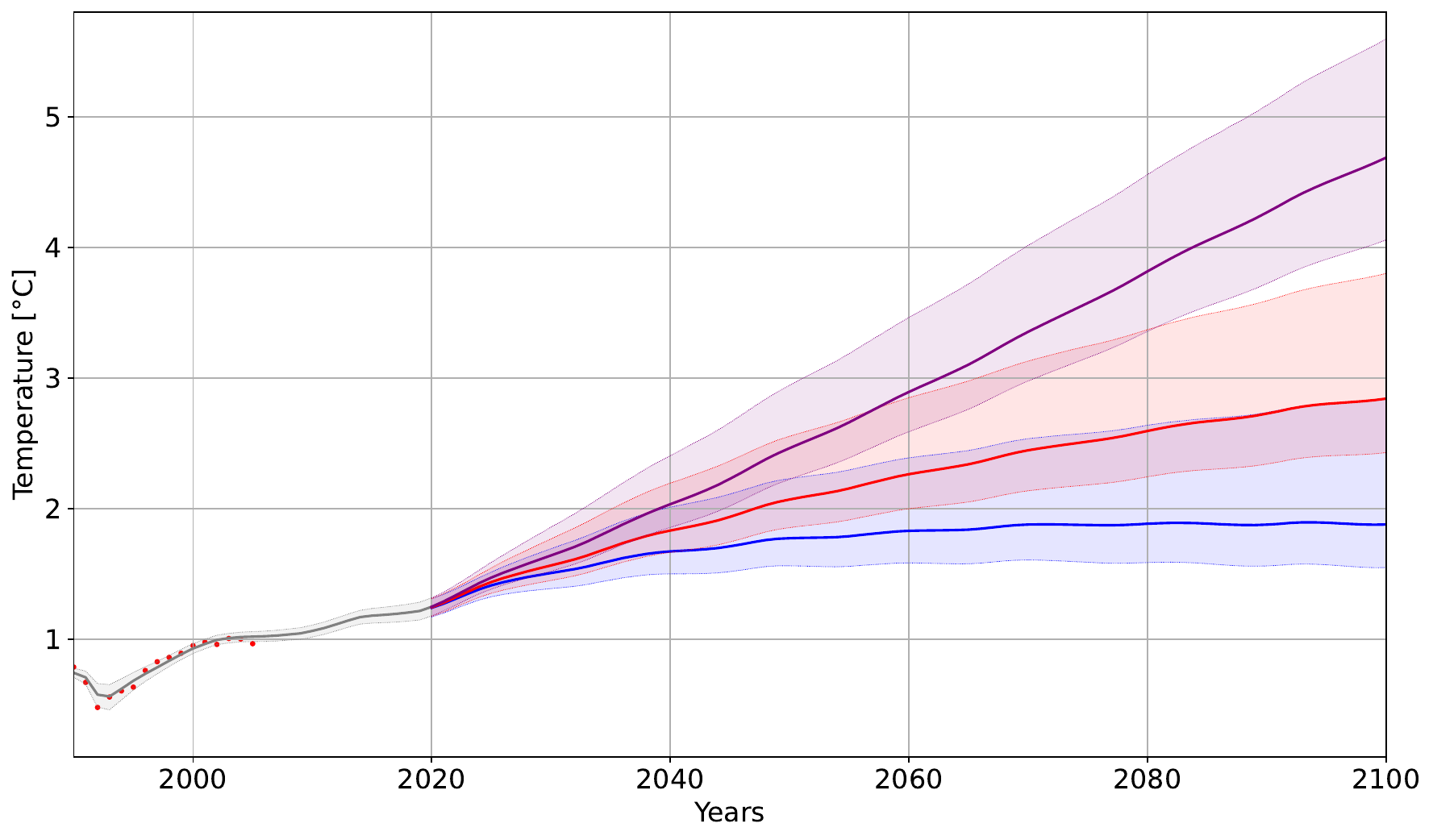}
    \captionsetup{font=footnotesize}
    \caption{Parameter uncertainty quantification and emission uncertainty quantification with 90\% confidence interval, FaIR is run until 2020 with same emissions and then scenarios are applied from there. Red dots represent historical data, Blue SSP1-RCP2.6, Red SSP2-RCP4.5, Purple SSP5-RCP8.5.}
    \label{fig:full_uq_all_scens}
\end{figure}

This test case integrates the methodologies from Test Cases 1 and 2 by combining parameter uncertainty quantification with emission uncertainty quantification. Specifically, we sample from two distributions: the emission uncertainty distribution described in Test Case 2 and the posterior distributions of climate model parameters obtained from Test Case 1. We use the parameters $\mu = 1$ and $\sigma = 13$ for a lognormal distribution. The results for the SSP1-RCP1.9, SSP2-RCP4.5, and SSP5-RCP8.5 scenarios, each with a 90\% confidence interval are presented in Figure \ref{fig:full_uq_all_scens}. 

\subsection{Test case 4: Real-world example SSAB}

We examine SSAB AB, Sweden’s leading steel producer, within the context of the socio-economic model outlined in Section \ref{subsubsec:soc-eco-model}. According to SSAB's 2023 annual report, \lq\lq There was increased interest in products with no carbon dioxide emissions during 2023 and we delivered more than 50,000 tonnes of SSAB Zero, a steel without Scope 1 and 2 emissions. Interest increased strongly not only in Europe, but also in the USA. We have started the construction of the electric arc furnace (EAF) in Oxel\"osund, a key item for fossil-free steelmaking. At the end of January 2024, the ruling on the granted concession for the power lines to Oxel\"osund gained legal force and the project continues according to plan.\rq\rq\ 
This report provides a pertinent example for testing our methodology. Specifically, we aim to assess the impact of this innovation through a hypothetical scenario: What if the entire sector operated with the reduced emission intensity of this new process?

Scope 1, 2 and 3 emissions for the year 2022 were obtained from SSAB's 2023 annual report \cite{ssab_anual_2023}. Gross value added was estimated using the income approach, which involves summing the following data points extracted from the annual report: operating profit, employee compensation, depreciation and amortization. This methodology enables the calculation of SSAB's economic emission intensity (EEI), which serves as an input to our model. The input data is summarized in Table \ref{tab:SSAB_Input} below. Note that the financial data points were originally reported in SEK and were converted to USD using the mean opening and closing exchange rate from \cite{investin_com_usd_sek} as of the year 2022.

\begin{table}[h]
    \centering
    \begin{tabular}{c|c}
        \hline
        Key figure & 2022 \\
        \hline
         Operating profit & 1,632.7 Mn USD\\
         Depreciation and Amortization & 364 Mn USD \\
         Employee compensation & 1,286 Mn USD \\
         Gross Value Added & 3,283 Mn USD \\
         Scope 1 Emission & 9,582 Gt $\text{CO}_2$e\\
         Scope 2 Emission & 1,179  Gt $\text{CO}_2$e\\
         Scope 3 Emission & 11,352 Gt $\text{CO}_2$e\\
         \hline
    \end{tabular}
    \captionsetup{font=footnotesize}
    \caption{Input values for SSAB}
    \label{tab:SSAB_Input}
\end{table}
As explained in Subsection \ref{subsubsec:soc-eco-model}, the first step in the socio-economic model is to calculate the global emissions \begin{equation} \label{eq:E_port_sect}
  \text{EMISSIONS}^{\text{Portfolio,Sector}} = \text{EEI}^{\text{Portfolio, Sector}}\text{GVA}^{\text{Sector}}
\end{equation} 
that correspond to the portfolios sectors' economic emission intensities. Since our portfolio only contains one constituent in the iron and steel sector, we substitute the sector emissions $\text{EMISSIONS}^{\text{iron and steel}}$ with $\text{EMISSIONS}^{\text{Portfolio, iron and steel}}$. See Section \ref{sec:sectors} in the Appendix for the full sector specification. 

Due to limited availability of sector-specific GVA data, we approximate \linebreak $\text{EMISSIONS}^{\text{Portfolio,Sector}}$ by utilizing a benchmark ensemble of companies within the same sector. This approach shifts the model towards a benchmark comparison framework. If the benchmark ensemble included every company in the sector worldwide, it would accurately represent the the sector's total GVA. The ensemble is employed to estimate a weighted mean of emission intensities, with weights based on each company's GVA. 
This approximation not only addresses the scarcity of sector-specific GVA data but also facilitates the transition from reported $\text{CO}_2$e data to encompass all physically relevant greenhouse gases included in the FaIR model. For a detailed explanation of this methodology, see Section \ref{seq:approx_EEI} in the Appendix. 

\begin{figure}[h]
    \centering
    \begin{minipage}{0.49\textwidth}
        \centering
        \includegraphics[height=4.4cm]{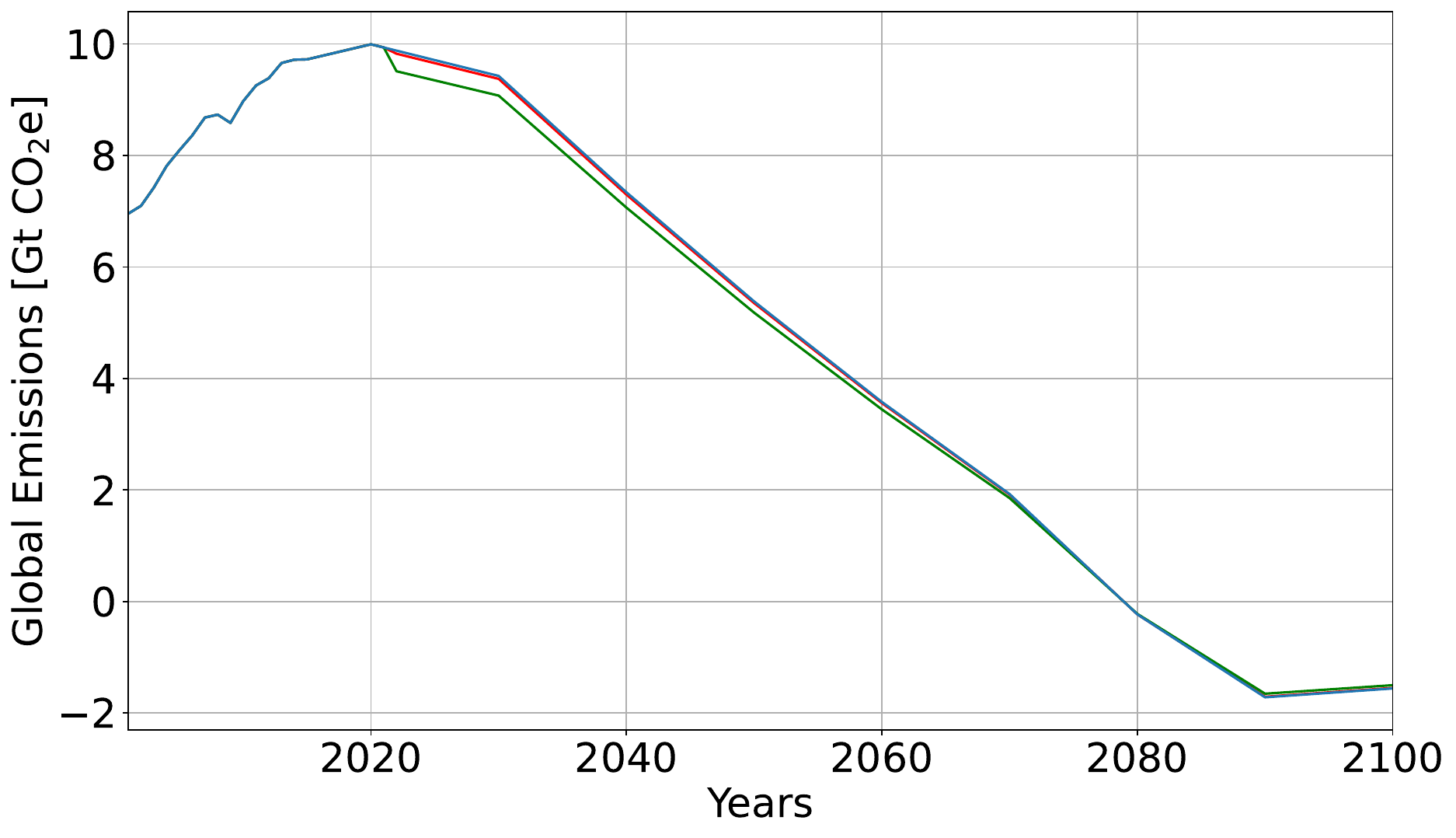}
        \captionsetup{font=footnotesize, margin=6pt}
        \centering
        \caption{Global emission by the socio-economic model in the SSP1-RCP2.6 scenario. Blue: SSP scenario, Red: current steel production, Green: green steel production.}
        \label{fig:globalEmissionsStoxxBenchmark126}
    \end{minipage}
    \hfill
    \begin{minipage}{0.49\textwidth}
        \centering
        \includegraphics[height=4.4cm]{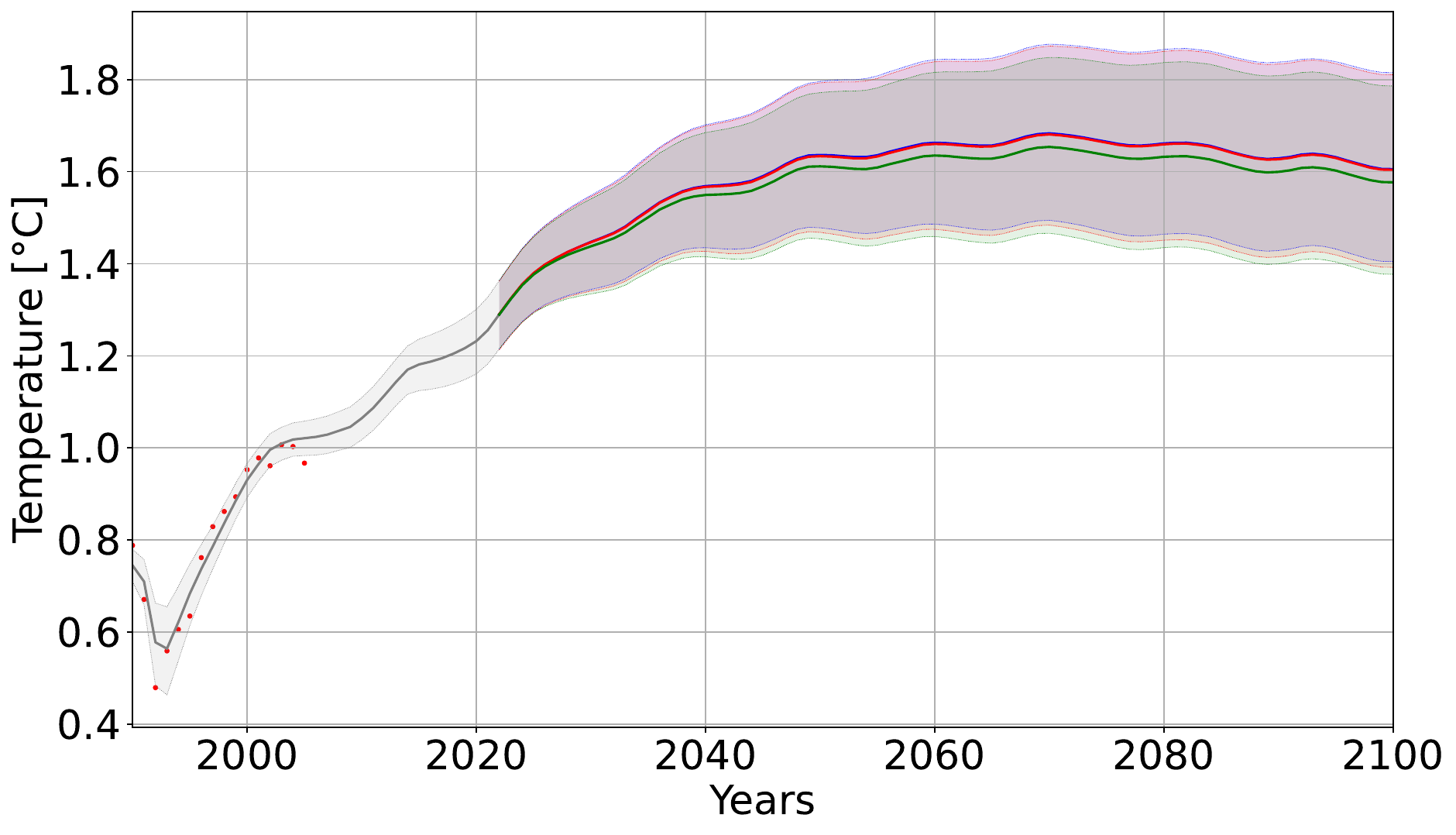}
        \captionsetup{font=footnotesize, margin=15pt}
        \centering
        \caption{Global temperature by the socio-economic model in the SSP1-RCP2.6 scenario. Blue: SSP scenario, Red: current steel production, Green: green steel.}
        \label{fig:globalTempStoxxBenchmark126}    
    \end{minipage}
\end{figure}

\begin{figure}[h]
    \begin{minipage}{0.49\textwidth}
        \centering
        \includegraphics[height=4.4cm]{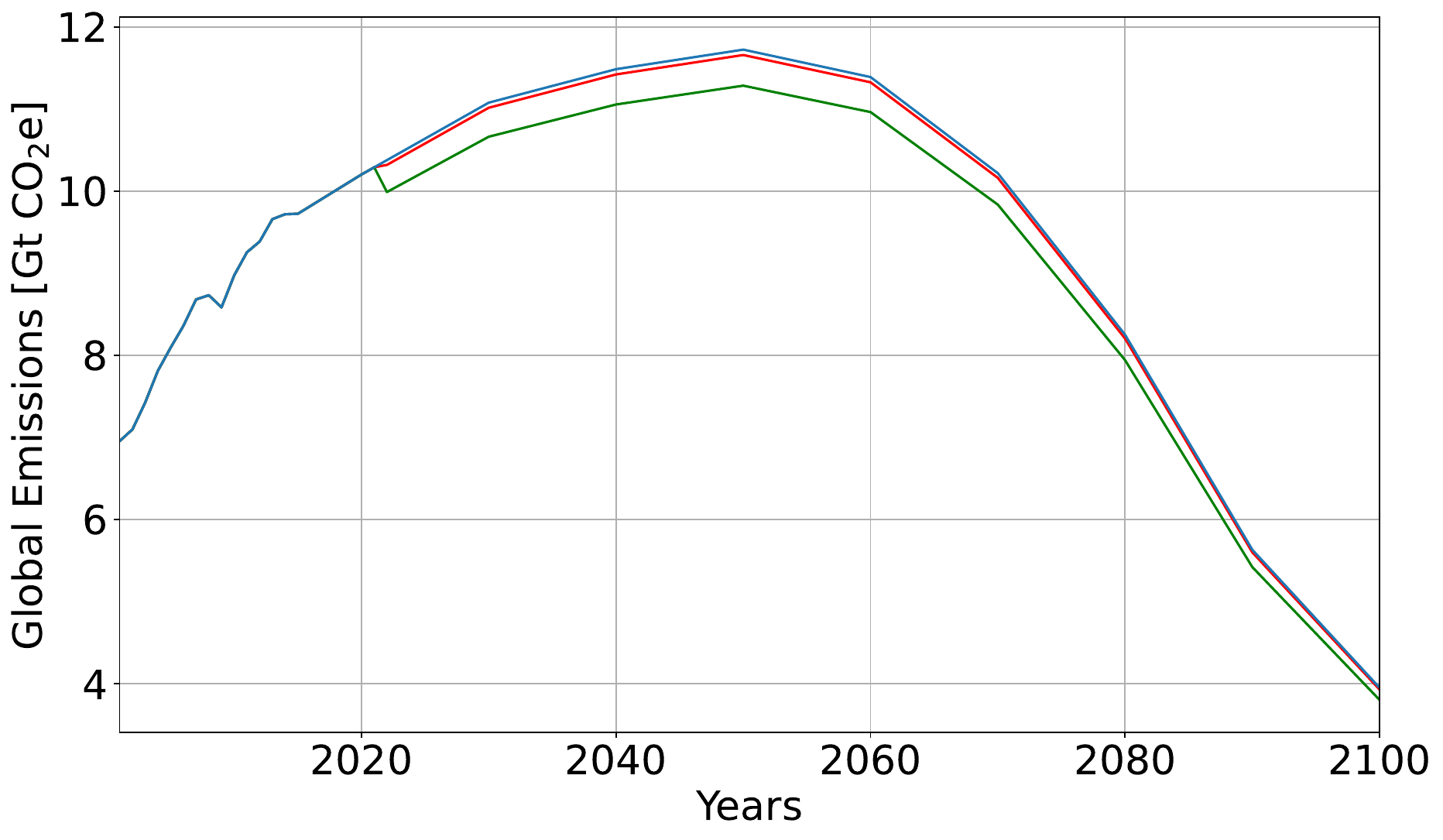}
        \captionsetup{font=footnotesize, margin=6pt}
        \centering
        \caption{Global emission by the socio-economic model in the SSP2-RCP4.5 scenario. Blue: SSP scenario, Red: current steel production, Green: green steel production.}
        \label{fig:global_emissions_stoxx_benchmark_245}
    \end{minipage}
    \hfill
    \begin{minipage}{0.49\textwidth}
        \centering
        \includegraphics[height=4.4cm]{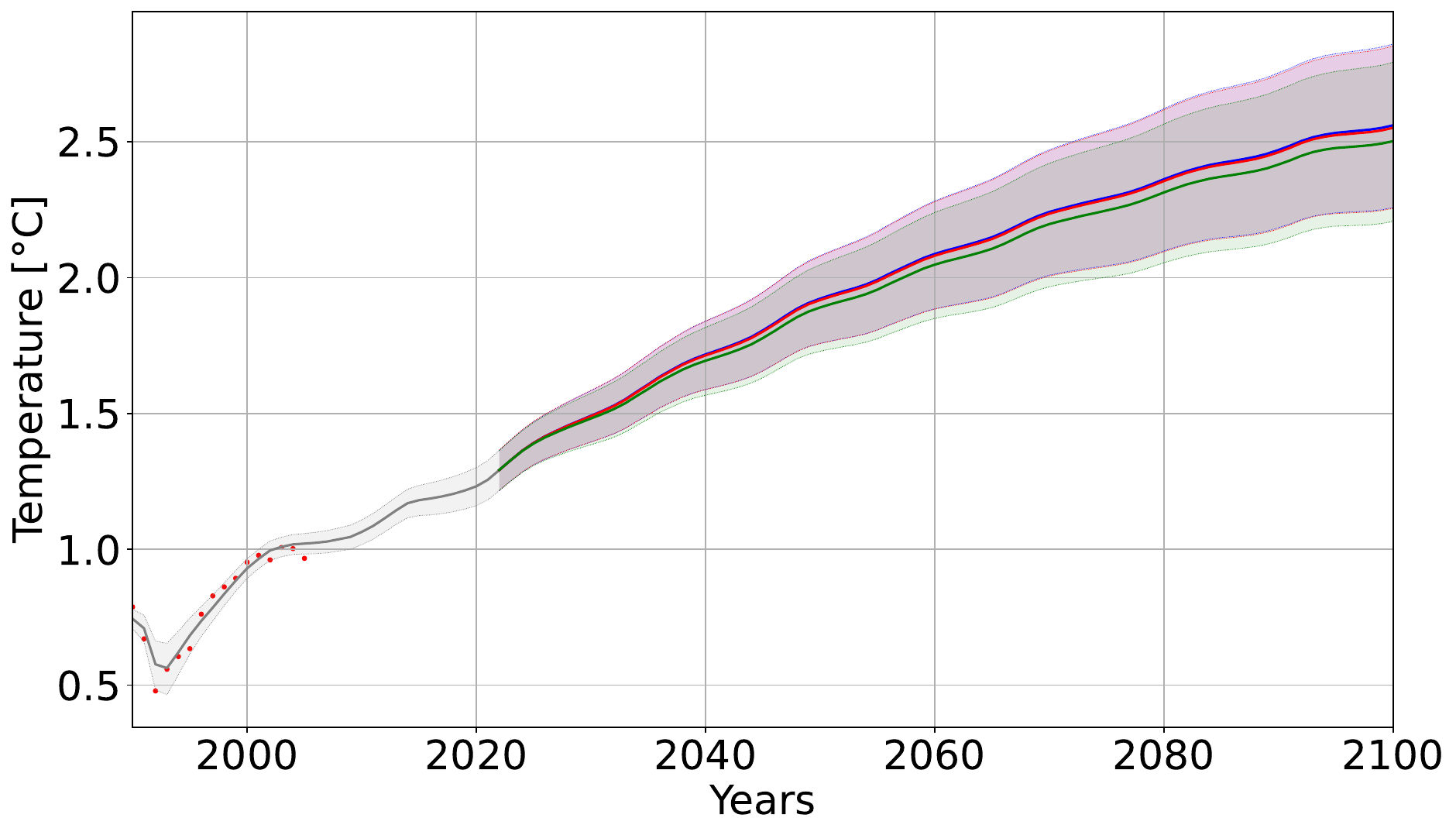} 
        \captionsetup{font=footnotesize, margin=15pt}
        \centering
        \caption{Global temperature by the socio-economic model in the SSP2-RCP4.5 scenario. Blue: SSP scenario, Red: current steel production, Green: green steel.}
        \label{fig:global_temp_stoxx_benchmark_245}    
    \end{minipage}
\end{figure}

\begin{figure}[h]
    \begin{minipage}{0.49\textwidth}
        \centering
        \includegraphics[height=4.4cm]{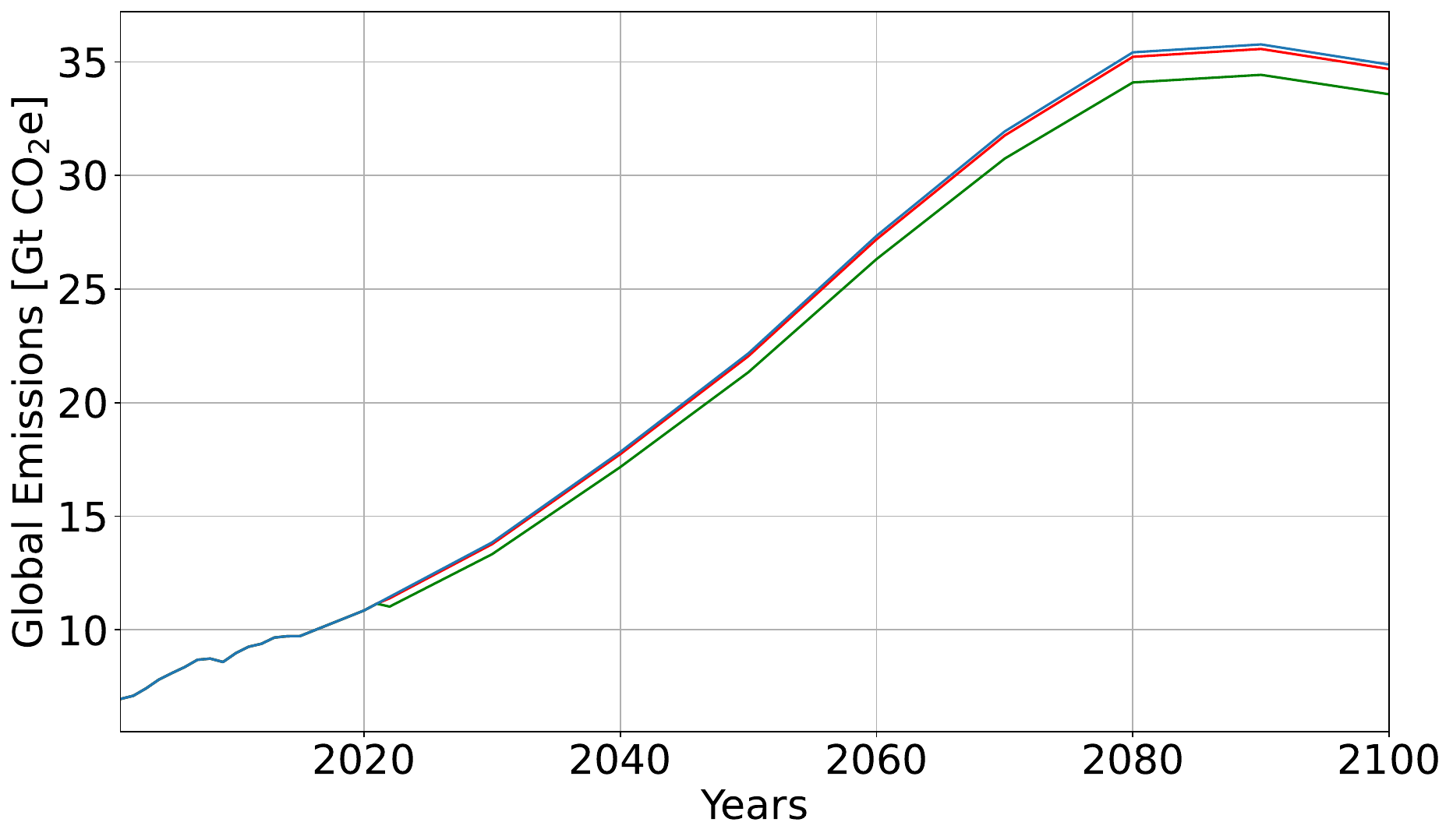}
        \captionsetup{font=footnotesize, margin=6pt}
        \centering
        \caption{Global emission by the socio-economic model in the SSP5-RCP8.5 scenario. Blue: SSP scenario, Red: current steel production, Green: green steel production.}
        \label{fig:global_emissions_stoxx_benchmark_585}
    \end{minipage}
    \hfill
    \begin{minipage}{0.49\textwidth}
        \centering
        \includegraphics[height=4.4cm]{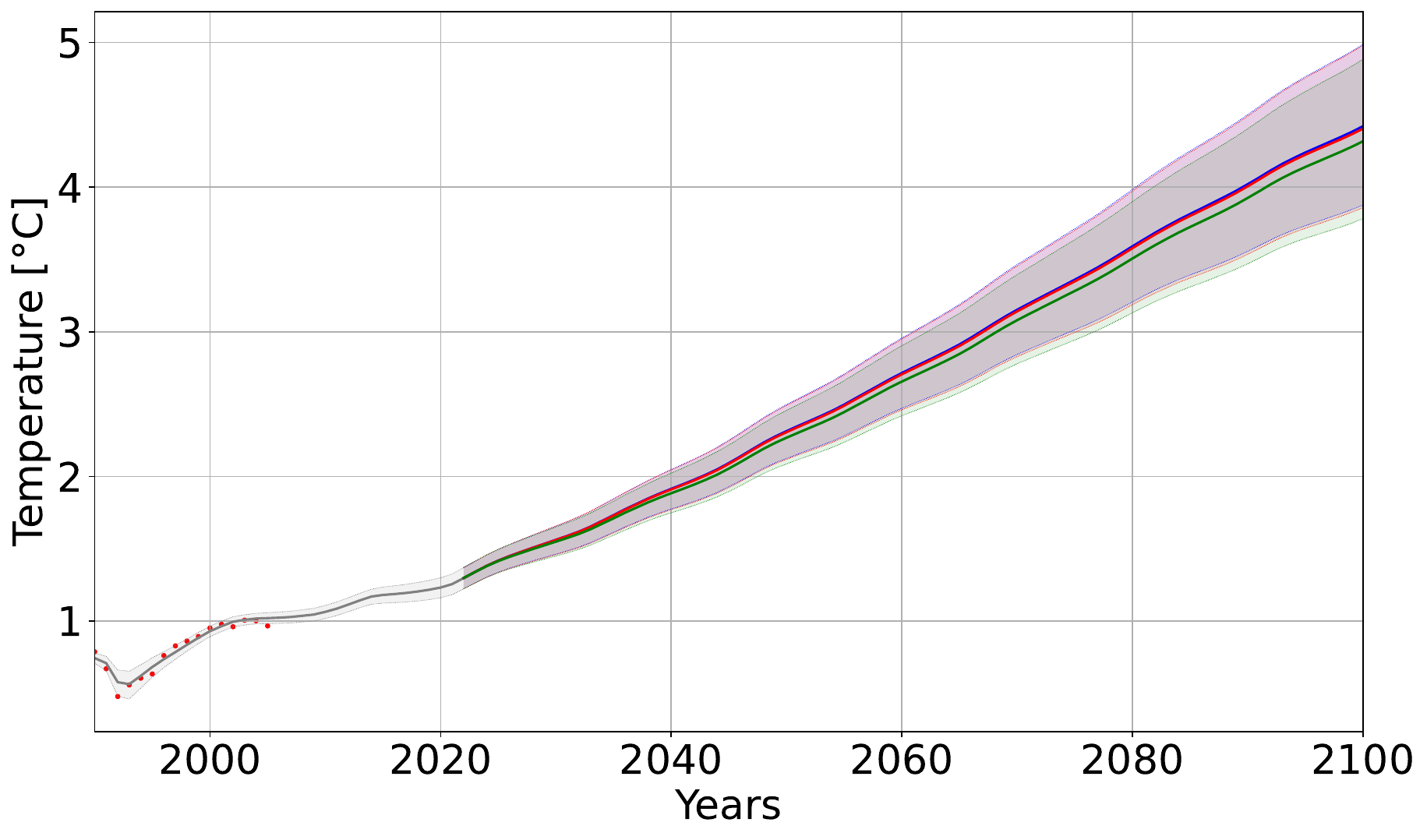}
        \captionsetup{font=footnotesize, margin=15pt}
        \centering
        \caption{Global temperature by the socio-economic model in the SSP5-RCP8.5 scenario. Blue: SSP scenario, Red: current steel production, Green: green steel.}
        \label{fig:global_temp_stoxx_benchmark_585}    
    \end{minipage}
\end{figure}

The novel green steel production method discussed in the report reduces the emissions from steel production from 2.4 kg $\text{CO}_2$e per kg steel to less than 0.05 kg $\text{CO}_2$e per kg steel \cite{ssab_green_steel}. In 2022, SSAB's total greenhouse gas emissions were reported at 22,375 kt \cite{ssab_anual_2023}. Based on our calculations, applying this reduction to all steel production-related emissions would result in a new total greenhouse gas emission level of approximately 11,570 kt. In this example, we demonstrate how the socio-economic model introduced in Section \ref{subsubsec:soc-eco-model} can be utilized to evaluate the impact of this emissions reduction on the Earth's temperature response. As detailed in Section \ref{seq:approx_EEI} in the Appendix, the global emissions approximation involves comparing the observed portfolio’s EEI to that of a benchmark ensemble. This process yields an estimated EEI for the sector. For this analysis, we use the STOXX Europe 600 stock index as the benchmark ensemble. To evaluate the global emissions impact, we focus on companies within the same sector as SSAB AB, namely iron and steel, and calculate the weighted mean of their EEIs. In 2022, the STOXX Europe 600 included four companies in the iron and steel sector: ArcelorMittal SA, thyssenkrupp AG, voestalpine AG, and SSAB AB. Using equation (\ref{eq:global_emissions}), we calculate the corresponding global emissions and analyze the results under three SSP scenarios. Figures \ref{fig:globalEmissionsStoxxBenchmark126}, \ref{fig:globalTempStoxxBenchmark126}, \ref{fig:global_emissions_stoxx_benchmark_245}, \ref{fig:global_temp_stoxx_benchmark_245}, \ref{fig:global_emissions_stoxx_benchmark_585}, and \ref{fig:global_temp_stoxx_benchmark_585} illustrate the global emissions pathways and the Earth's temperature response for each respective SSP scenario: SSP1-RCP2.6, SSP2-RCP4.5, SSP5-RCP8.5. 

The approximated sector EEI of the STOXX Europe 600 benchmark is 5,183.11 \linebreak t$\text{CO}_2$e/Mn USD. According to SSAB's 2022 annual report, its EEI is approximately 4,902.0 t$\text{CO}_2$e/Mn USD. With the adoption of the newly implemented steel production method and assuming a constant GVA, SSAB's EEI decreases to 2,564.9 t$\text{CO}_2$e/Mn USD. Table \ref{tab:temp_output} presents the resulting temperature outcomes based on these EEIs. Within the SSP2-RCP4.5 scenario, this reduction corresponds to a temperature improvement from the baseline of 2.558°C to 2.551°C. With the new steel production method, the aligned temperature is 2.501°C. These temperature values represent the mean outcomes from the Monte Carlo simulations described in Subsection \ref{subsec:test_emissions_unc}, i.e. with parameter uncertainty quantification.

\begin{table}[h]
    \scriptsize
    \centering
        \begin{tabular}{c|c c c }
         \hline
         Scenario & Baseline Temp. & Temp. SSAB & Temp. of SSAB with green steel \\
         \hline
         SSP1-RCP2.6 & 1.611°C & 1.604°C & 1.584°C \\
         SSP2-RCP4.5 & 2.558°C & 2.551°C & 2.501°C \\
         SSP5-RCP8.5 & 4.444°C & 4.433°C & 4.327°C\\
         \hline
    \end{tabular} 
    \captionsetup{font=scriptsize}
    \caption{Temperatures in SSP-scenarios for SSAB test case}
    \label{tab:temp_output}
\end{table}

\subsection{Discussion}
The first test case highlights the advantages of using Bayesian parameter calibration over simpler sensitivity-based methods. The results demonstrate that Markov Chain Monte Carlo (MCMC) simulations applied to FaIR's parameter uncertainty significantly narrow the confidence intervals of temperature forecasts. This improvement underscores the value of Bayesian methods in enhancing forecast reliability and reducing uncertainty in long-term climate modeling. The proposed approach effectively reduces computational intensity through the use of deep learning techniques.

By incorporating emission uncertainties via PDFs, the second test case addresses the critical uncertainties surrounding reported emissions, particularly in light of concerns about carbon offsets. The choice between lognormal and normal distributions demonstrates flexibility in modeling systematic biases or random deviations. The lognormal distribution, suited for cases like underreporting, provides a tailored approach for real-world scenarios. The outcomes highlight how incorporating such uncertainties refines emission pathways and temperature forecasts. The findings suggest that future research should focus on refining distribution parameters, leveraging resources like the EDGAR database to improve precision. This approach effectively bridges scientific rigor and practical policy needs but requires additional validation across diverse datasets.

The third test case integrates the methodologies from the first two cases, combining parameter and emission uncertainties. The results reveal how these combined sources of uncertainty propagate through climate models and influence temperature pathways under various SSP scenarios. This combined approach provides a more realistic representation of real-world variability, making it particularly useful for policy analysis. 

The fourth test case illustrates the practical application of the framework using SSAB AB, a leading steel producer transitioning to low-emission steel production. The socio-economic model effectively quantifies the temperature impact of reduced emissions, showing how SSAB’s adoption of green steel production methods could lower its EEI. In the SSP2-RCP4.5 scenario, this transition results in a temperature improvement from 2.56°C to 2.50°C, demonstrating the tangible benefits of industry-wide emission reductions. While the case highlights the potential for transformation within the sector, it underscores that a successful transition requires collaborative efforts across all sectors. At the same time, our framework offers a scientifically rigorous basis for comparing the transition risks faced by different companies within the sector.
This test case also highlights data challenges, such as approximating global emissions based on limited GVA data, which underscores the need for both, further transparency and research.

%%%%%%%%%%%%%%%%%%%%%%%%%%%%%%%%%%%%%%%%%%%%%%%%%%%%%%%%%%%%%%%%%%%%%%%%%%%%%%%%%%%%%%%%%%%%

\section{Conclusion}
Using temperature alignment with uncertainty quantification allows investors to integrate climate considerations into portfolio construction in a scientifically rigorous and decision-relevant manner. Uncertainty quantification captures the variability and unreliability inherent in emissions data, climate models, and climate model parameters, making it easier to compare companies and portfolios in terms of their long-term climate impact. By incorporating these uncertainties, investors can better assess the robustness of their portfolios under different climate scenarios.

The framework presented in this work not only enhances the credibility of climate-aligned investment strategies but also provides a practical tool for identifying transition risks and opportunities. By linking emissions data and decarbonization pathways to global temperature outcomes, it ensures that investment decisions are aligned with broader climate goals, such as limiting global warming to 1.5°C or 2°C. While the methodology offers significant advancements, we also highlight areas for future research. These include refining uncertainty modeling, validating the framework across diverse sectors and geographies, and exploring the implications of systemic interactions between sectors. Addressing these challenges will further enhance the reliability and applicability of the approach. Exploring model uncertainty by comparing different versions of the FaIR model with other simple climate models, such as MAGICC, cf. \cite{meinshausen2020shared} and the widely used linear TCRE-based solutions, is an equally compelling avenue for future research. Finally, embedding the framework directly into portfolio construction processes for climate investing, see, e.g., \cite{le2022portfolio}, offers a highly impactful and promising direction for future research. Integrating temperature alignment and uncertainty quantification directly into optimization algorithms has the potential to enable investors to design portfolios that are not only aligned with a climate target but also resilient to data and model uncertainties. This will further enhance the practical relevance of the framework and expand its adoption in the financial sector. In a follow-up publication, we aim to investigate portfolio optimization that seamlessly integrates the output of our temperature alignment framework, along with uncertainty quantification, directly into the objective function.

In conclusion, temperature alignment with uncertainty quantification represents a critical step forward in aligning investment strategies with global climate objectives. By equipping investors with robust, transparent, and actionable insights, this approach supports informed decision-making and fosters the transition to a low-carbon economy.

%%%%%%%%%%%%%%%%%%%%%%%%%%%%%%%%%%%%%%%%%%%%%%%%%%%%%%%%%%%%%%%%%%%%%%%%%%%%%%%%%%%%%%%%%%%%

\pagebreak
\appendix

\section{Shared Socioeconomic Pathways (SSP) narratives}
Five Shared Socioeconomic Pathways (SSP) narratives were initially developed to describe different societal developments\footnote{We closely follow the descriptions provided by the Deutsches Klimarechenzentrum (DKRZ) at \url{https://www.dkrz.de/en/communication/climate-simulations/cmip6-en/the-ssp-scenarios}}:

\textbf{SSP1}: The sustainable and green path describes an increasingly sustainable world. Global common goods are preserved, and the limits of nature are respected. Human well-being becomes the focus instead of economic growth. Income inequalities between and within countries are reduced, and consumption is oriented toward low material and energy use.

\textbf{SSP2}: The middle-of-the-road scenario continues current development trends. Income developments of individual countries vary widely. There is some cooperation between states, but it progresses only slightly. Global population growth is moderate and slows down in the second half of the century. Environmental systems experience some degradation.

\textbf{SSP3}: Regional Rivalries. Nationalism and regional conflicts re-emerge, pushing global issues to the background. Policies increasingly focus on national and regional security concerns. Investments in education and technological development decline. Inequalities increase, and some regions experience severe environmental degradation.

\textbf{SSP4}: Inequality. The gap widens between developed societies, which also cooperate globally, and those stuck at low development levels with low income and education. In some regions, environmental policies are successful at addressing local problems, while in others, they are not.

\textbf{SSP5}: Fossil-fueled development. Global markets become increasingly integrated, leading to innovations and technological advancements. However, social and economic development is based on the intensified exploitation of fossil fuel resources with a high coal share and an energy-intensive lifestyle worldwide. The global economy grows, and local environmental issues like air pollution are successfully addressed.

%%%%%%%%%%%%%%%%%%%%%%%%%%%%%%%%%%%%%%%%%%%%%%%%%%%%%%%%%%%%%%%%%%%%%%%%%%%%%%%%%%%%%%%%%%%%

\section{The Bayesian posterior density} \label{ap:bayesian_posterior}
\begin{figure}[H]
    \centering
    \includegraphics[width=\textwidth]{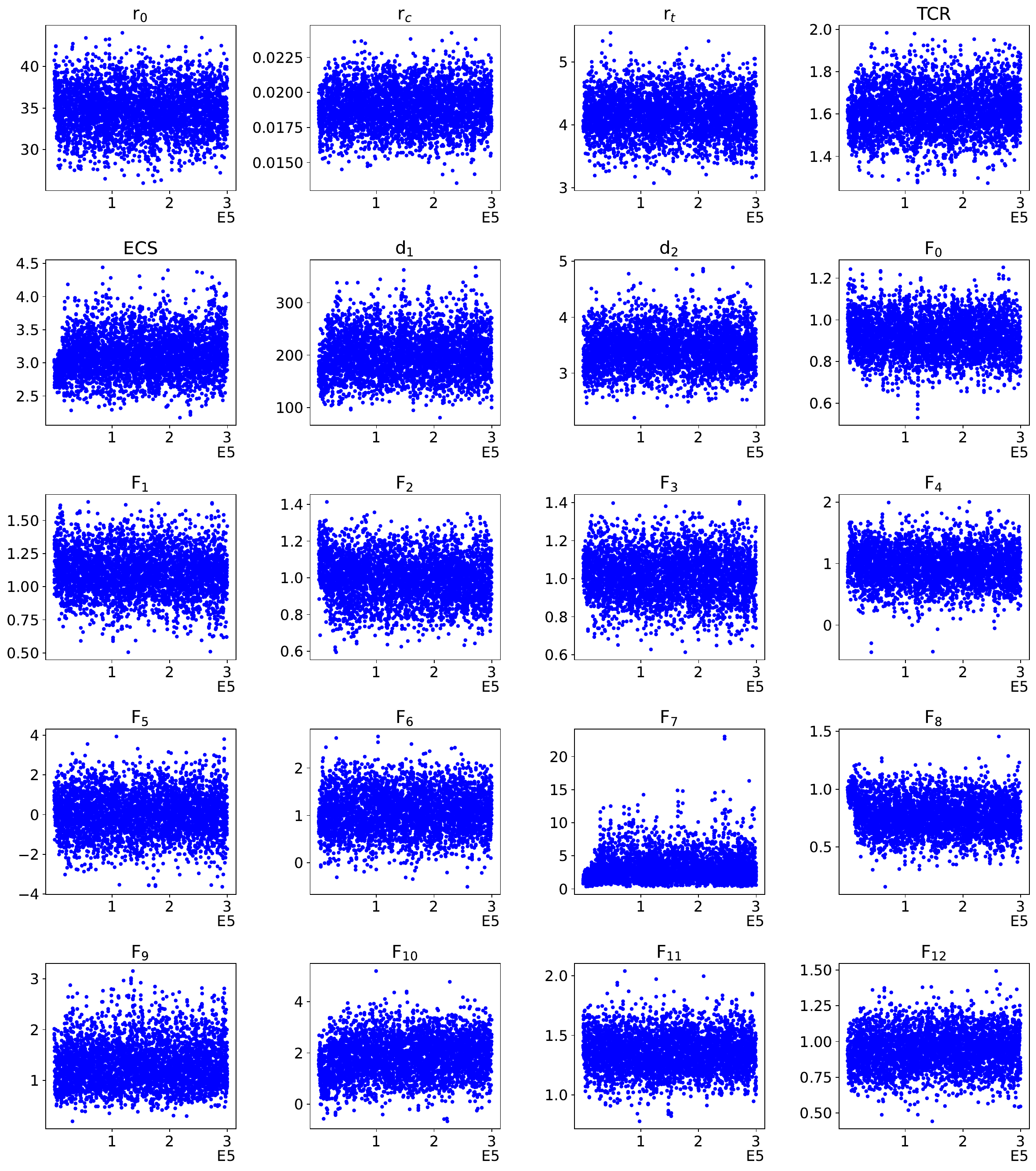}
    \captionsetup{font=footnotesize}
    \caption{Chains sampled from the posterior corresponding to the statistical calibration problem for the FaIR model obtained from sampling with DRAM.}
    \label{fig:chains}
\end{figure}

The posterior distribution of the unknown parameter vector $\boldsymbol{\theta}$ given the historical temperature observations $\boldsymbol{y}$ is given by Bayes' formula
\begin{equation}\label{eqn:posterior}
p(\boldsymbol{\theta}|\textbf{y})=\frac{p(\textbf{y}|\boldsymbol{\theta})p(\boldsymbol{\theta})}{\int p(\textbf{y}|\boldsymbol{\theta})p(\boldsymbol{\theta}) \mathrm{d}\boldsymbol{\theta}},
\end{equation}
where $p(\boldsymbol{\theta}|\textbf{y})$ denotes the likelihood of 
$\boldsymbol{\theta}$ for given data $\textbf{y}$, 
$p(\textbf{y}|\boldsymbol{\theta})$ gives likelihood of $\textbf{y}$ for given $\boldsymbol{\theta}$, and 
$p(\boldsymbol{\theta})$ denotes the prior distribution of $\boldsymbol{\theta}$. Bayes formula provides a framework to solve the estimation problem, though practical applications present several challenges. Defining the prior for the parameters typically requires expert knowledge; in this work, we rely on the assumptions used in reference \cite{smith2018fair}. A primary challenge in a straightforward application of Bayes' formula is calculating the integral in the denominator of equation (\ref{eqn:posterior}), which, in our case, requires a 20-dimensional integration. Since this is practically infeasible, we rely on adaptive MCMC sampling methods to explore the posterior distribution.
Figure \ref{fig:chains}  illustrates an example of the chains produced via MCMC sampling, showing the sampled values for each of the 20 FaIR parameters as a function of the iteration count. We observe that, after an initial adaptation period, the sampled values reach stationary behavior, indicating parameter values that make the model align with the observed historical temperature data within the specified error bounds.

%%%%%%%%%%%%%%%%%%%%%%%%%%%%%%%%%%%%%%%%%%%%%%%%%%%%%%%%%%%%%%%%%%%%%%%%%%%%%%%%%%%%%%%%%%%%

\section{Origin of sector-wise emission data}\label{sec:sectors}
The socio-economic model relies heavily on sector-specific emission pathways, which are embedded within climate scenarios. In this study, we adapt the SSP scenarios to derive sector-specific pathways. The International Energy Agency (IEA) presents the Net Zero Scenario, which includes sector-wise emission pathways \cite{iea2021netzero}. These pathways outline an optimistic trajectory toward a net-zero emission economy by 2050, with emission reductions of approximately 36\% between 2020 and 2030, and about 72\% between 2030 and 2050.
The IEA categorizes emissions into the sectors and subsectors depicted in Table \ref{tab:sectors_iea}.

\begin{table}[h]
    \centering
    \scriptsize
    \renewcommand{\arraystretch}{1.2}  
    \begin{tabular}{l|c}
         \Xhline{2\arrayrulewidth}
         Sector & Share of global emissions \\
         \hline
         Industry & 25.01\% \\
         \quad Iron and steel & 6.93\% \\
         \quad Chemicals & 3.82\% \\
         \quad Cement & 6.88\% \\
         \hline
         Transport & 21.10\% \\
         \quad Road and passenger cars & 8.10\% \\
         \quad Road trucks & 5.08\% \\
         \quad Aviation & 1.83\% \\
         \quad Shipping & 2.36\% \\
         \hline
         Buildings & 8.44\% \\
         \quad Residential & 5.80\% \\
         \quad Services & 2.88\% \\
         \hline
         Electricity and Heat & 44.17\% \\
         \Xhline{2\arrayrulewidth}
    \end{tabular}  
    \caption{Sector-wise global emissions in 2020 \cite{iea2021netzero}}
    \label{tab:sectors_iea}
\end{table}

To perform a comprehensive climate analysis, it is essential to consider a wide range of scenarios. As discussed in Section \ref{sec:scenario_uncertainty}, SSP scenarios odder a diverse set of pathways and are extensively utilized both in academia and practice. However, since SSP scenarios do not provide sector-specific emission data, we leverage the sector-wise allocation percentages from the the IEA's Net Zero Scenario to disaggregate the global emission pathways into sector-specific pathways.

%%%%%%%%%%%%%%%%%%%%%%%%%%%%%%%%%%%%%%%%%%%%%%%%%%%%%%%%%%%%%%%%%%%%%%%%%%%%%%%%%%%%%%%%%%%%

\section{Approximation of the sector emissions corresponding to a portfolio} \label{seq:approx_EEI}
The sector-specific global gross emissions, $\text{EMISSIONS}^{\text{Sector}}$, is a parameter in the socio-economic model with robust data availability and quality; it can be obtained, e.g., from scenarios given in the Net Zero Scenario by the IEA \cite{iea2021netzero}. % TODO referece origin of sectoral emission data 
It can be expressed in terms of the quantities $\text{EEI}^{\text{Sector}}$ and $\text{GVA}^{\text{Sector}}$, which are generally unknown:
\begin{equation} \label{eq:sector_emission}
      \text{EMISSIONS}^{\text{Sector}} =  
      \text{EEI}^{\text{Sector}}\text{GVA}^{\text{Sector}}.
\end{equation}
To calculate $\text{EMISSIONS}^{\text{Portfolio,Sector}}$, which represents the global emissions aligned with the portfolio's economic emission intensity, we substitute the sector emission intensity $\text{EEI}^{\text{Sector}}$ with the portfolio's emission intensity for this specific sector, $\text{EEI}^{\text{Portfolio, Sector}}$ (see equation (\ref{eq:E_port_sect})). To utilize a a proxy emission intensity, $\widehat{\text{EEI}}^{\text{Sector}}$, and incorporate a benchmark comparison, we adapt equation (\ref{eq:E_port_sect}) as follows: 
\begin{equation*}
      \text{EMISSIONS}^{\text{Portfolio,Sector}} =  
    \frac{\text{EEI}^{\text{Sector}}}{\widehat{\text{EEI}}^{\text{Sector}}}
      \text{EEI}^{\text{Portfolio, Sector}}\text{GVA}^{\text{Sector}}.
\end{equation*}
From (\ref{eq:sector_emission}), we obtain the sector-specific emissions corresponding to the portfolio as follows:
\begin{equation} \label{eq:emissions_P_S}
      \text{EMISSIONS}^{\text{Portfolio,Sector}} =  
    \frac{\text{EEI}^{\text{Portfolio, Sector}}}{\widehat{\text{EEI}}^{\text{Sector}}}
      \text{EMISSIONS}^{\text{Sector}}.
\end{equation}
As explained in Section \ref{subsubsec:soc-eco-model}, $\text{EMISSIONS}^{\text{Portfolio,Sector}}$ is calculated for all sectors that are represented in the portfolio. For the sectors that are not represented in the portfolio, we stick with the original data for the sector. The present-day, global emission are the sum of all sector-emissions, i.e.: 
\begin{equation} \label{ap:eq:global_emissions}
    \text{EMISSION}^{\text{Global}}=\sum_{\text{Sectors}} \text{EMISSIONS}^{\text{Portfolio,Sector}}
\end{equation}
we then apply the growth curve, of a SSP-sceanrio to obtain a pathway of emissions from the base year until 2100. With these emission a temperature alignment is carried out through the FaIR climate model described in \ref{subsubsec:climate_model}.
\\ \\
The FaIR climate model can operate with both a one-dimensional input array of $\text{CO}_2$-equivalent emissions, and with a more comprehensive two-dimensional input array of 39 greenhouse gases. Our analysis indicates that incorporating the complete set of greenhouse gases in FaIR’s retrospective simulations yields results more closely aligned with recorded temperature data; despite the fact that most companies only publish their emission in $\text{CO}_2$e, we make use of the more precise multigas mode with the following procedure. We expand the equation (\ref{eq:emissions_P_S}) through calculating sector emission for each greenhouse gas as follows: 
\begin{equation} \label{eq:global_emissions}
      \text{EMISSIONS}^{\text{Portfolio,Sector}}_{\text{Gas}} =  
    \frac{\text{EEI}^{\text{Portfolio, Sector}}_{\text{CO}_2\text{e}}}{\widehat{\text{EEI}}^{\text{Sector}}_{\text{CO}_2\text{e}}}
      \text{EMISSIONS}^{\text{Sector}}_{\text{Gas}}.
\end{equation}
Similarly, we adapt equation (\ref{ap:eq:global_emissions}) such that we finally obtain an emission value for each greenhouse gas. The gases are among others $\text{CO}_2$, methane, nitrogen and sulphur dioxide; the full list of of greenhouse gases in the FaIR model can be seen in Table 1 in the original paper that introduces FaIR \cite{smith2018fair}.

%%%%%%%%%%%%%%%%%%%%%%%%%%%%%%%%%%%%%%%%%%%%%%%%%%%%%%%%%%%%%%%%%%%%%%%%%%%%%%%%%%%%%%%%%%%%

\section{Retrieval of data for the SSAB test case}

The data required for the benchmark was taken from annual reports. Since thyssenkrupp AG and voestalpine AG report their emissions in the fiscal year October 1st to September 30th and April 1st to March 30th, their emission and financial figures were calculated proportionately. The gross value added (GVA) is the sum of EBITDA and Employee compensation. For the conversion from SEK to USD and Euro to USD, we used $1 \text{ SEK} = 0.0988 \text{ USD}$ and $1 \text{ Euro} = 1.0516 \text{ USD}$ respectively. The conversion rate for SEK to USD are the mean average from all closing and opening rates in 2022 with the data from \cite{investin_com_usd_sek}, the conversion rates for Euro to USD are from \cite{euro_usd_conversion}. Since employee compensation was not publicily available for thyssenkrupp AG and voestalpine, the employee compensation was calculated as the product of average employee compensation from the other companies and the reported employee count in their annual reports \cite{thyssen_anual_2023, voestalpine_annual_2023}.

 \begin{table}[h]
    \centering
    \scriptsize
    \begin{threeparttable}
    \renewcommand{\arraystretch}{1.2}  
    \begin{tabular}{c|c c c c c c}
         \hline
         Company & GVA \tnote{1} & EBITDA \tnote{1}  & Employee compensation \tnote{1}  & Scope 1 \tnote{2} & Scope 2 \tnote{2} & Scope 3 \tnote{2}\\
         \hline
         ArcelorMittal SA & 19,354   & 14,161  & 5,193  & 112.9 & 6.1 & 6.1 \\
         thyssenkrupp AG & 8,149.17  & 2,121.1  & 6,028.07 & 22.525 & 0.95 & 3.9 \\
         voestalpine AG & 5,459.34  & 2,476.2  & 2,983.15 & 12.71 & 0.48 & 11.31 \\
         SSAB AB & 3,282  & 1,996  & 1,286  & 9.582 & 1.179 & 11.352 \\
         \hline
    
    \end{tabular}
    \begin{tablenotes}
        \item[1]{in Million USD.}
        \item[2]{in Mt $\text{CO}_2$e.}
    \end{tablenotes}        
    \captionsetup{font=scriptsize}
    \caption{Companies within STOXX 600 Europe 2022 that are participating in iron and steel sector and the required values for the socioeconomic model. All values are taken from their annual reports \cite{ssab_anual_2023, thyssen_anual_2023, voestalpine_annual_2023, arcelor_annual, thyssenkrupp_emission} unless stated otherwise.}
    
    \end{threeparttable}
\end{table}

\pagebreak
\section*{Acknowledgements}

The authors would like to express their gratitude to Emanuele Pepe and the team of right°, whose support and feedback significantly contributed to the development of this study. The content is solely the responsibility of the authors.

\section*{Funding}
This project (HA project no. 1647/23-200) is financed with funds of LOEWE – Landes-Offensive zur Entwicklung Wissenschaftlich-ökonomischer
Exzellenz, Förderlinie 3: KMU-Verbundvorhaben (State Offensive for the Development of Scientific and Economic Excellence). Heikki Haario's work has been supported by the Research Council of Finland (RCoF) through the Flagship of advanced mathematics for sensing, imaging and modelling, decision number 358 944.
Martin Simon would like to acknowledge support by the German Federal Ministry of Education and Research (BMBF) under Grant No. 03FHP191. 

%% References
\bibliographystyle{elsarticle-num}
\bibliography{bibliography}

\end{document}